\documentclass[a4paper,11pt]{article}
\usepackage{jheppub} 
\usepackage{lineno}
\usepackage{amsmath}
\usepackage{tikz}
\usetikzlibrary{decorations.pathmorphing, arrows.meta}
\usepackage{array}
\newcolumntype{L}[1]{>{\raggedright\arraybackslash}p{#1}}
\usepackage{comment}
\usepackage[x11names]{xcolor}
\makeatletter
\newcommand{\Romannum}[1]{\uppercase\expandafter{\romannumeral #1\relax}}
\makeatother
\usepackage{graphicx}
\usepackage[export]{adjustbox}
\graphicspath{ {./images/} }
\usepackage{hyperref}
\hypersetup{colorlinks=true, linkcolor=blue, filecolor=magenta, urlcolor=cyan,}
\usepackage{bbold}
\usepackage{tikz}
\usetikzlibrary{decorations.markings}
\usetikzlibrary{decorations.pathmorphing}
\usetikzlibrary{decorations.markings}
\usetikzlibrary{decorations.pathmorphing}
\usetikzlibrary{patterns.meta}
\usepackage[utf8]{inputenc}
    \usepackage{tikz}
    \usetikzlibrary{patterns}

\preprint{}
\title{Wilson coefficients from a non-renormalization theorem in  2D SYM}

\author[a]{Kabir Bajaj}
\affiliation[a]{Jadwin Hall, Princeton University,\\Princeton, NJ 08540, U.S.A.}
\emailAdd{kb6228@princeton.edu}

\abstract{Matrix string theory \cite{Dijkgraaf:1997vv,Motl:1997th,Seiberg:1997ad} is a conjectured duality between two-dimensional maximally supersymmetric $U(N)$ Yang--Mills theory and type-\Romannum{2}A string theory in ten-dimensional Minkowski spacetime. The IR description of this gauge theory is governed by the symmetric product orbifold \((\mathbb R^8)^N/S_N\) CFT. The leading irrelevant deformation from this IR fixed point is the Dijkgraaf--Verlinde--Verlinde (DVV) operator, which comes with an unknown Wilson coefficient. We determine this coefficient using non-renormalization arguments from the UV gauge theory. The result is consistent with the matrix string theory conjecture and gives a first-principles check of the relation between $g_{\rm{YM}}$ and the string coupling. We also comment on the prospects for fixing further Wilson coefficients using similar methods.}
\begin{document}
\maketitle
\flushbottom
\let\svthefootnote\thefootnote
\newcommand\blfootnotetext[1]{%
  \let\thefootnote\relax\footnote{#1}%
  \addtocounter{footnote}{-1}%
  \let\thefootnote\svthefootnote%
}

\let\svfootnotetext\footnotetext
\renewcommand\footnotetext[2][?]{%
  \if\relax#1\relax%
    \ifnum\value{footnote}=0\blfootnotetext{#2}\else\svfootnotetext{#2}\fi%
  \else%
    \if?#1\ifnum\value{footnote}=0\blfootnotetext{#2}\else\svfootnotetext{#2}\fi%
    \else\svfootnotetext[#1]{#2}\fi%
  \fi
}
\section{Introduction}
Matrix string theory (MST) \cite{Banks:1996my,Dijkgraaf:1997vv,Motl:1997th} is the conjecture that the target-space S-matrix of two-dimensional maximally supersymmetric $U(N)$ Yang--Mills theory compactified on a circle is equal to the S-matrix of type-\Romannum{2}A string theory in ten-dimensional Minkowski spacetime. In the IR limit of the gauge theory one recovers free Green--Schwarz light-cone strings, and the commuting Cartan sector of the gauge theory becomes the symmetric product orbifold \((\mathbb R^8)^N/S_N\), while string interactions are generated by the Dijkgraaf--Verlinde--Verlinde (DVV) twist-field deformation. The structure of this deformation is highly constrained by maximal supersymmetry in both the IR and UV, but the Wilson coefficient in front of it is not known a priori. Using this deformation from the orbifold side, highly nontrivial evidence for the proposal was provided in \cite{Arutyunov:1997gi,Arutyunov:1998eq}, where closed-string scattering amplitudes with external massless states were reproduced up to this undetermined Wilson coefficient.

We fix this coefficient by looking at the low-energy, large-impact-parameter limit of the amplitude. This limit is particularly well studied in the context of BFSS \cite{Becker:1997wh,Becker:1997xw,Banks:1996vh}\footnote{For a recent review see \cite{lin2025tasilecturesmatrixtheory,MaldaTalk}.}. Some indirect early evidence that the one-loop term in this limit is protected came from matching to the supergravity calculation \cite{Maldacena:1997kk}. The matrix-theory one-loop result gives the same numerical coefficient as the gravity computation.

For matrix quantum mechanics, non-renormalization theorems for the leading long-distance term were established directly from the gauge theory in \cite{Paban:1998ea,Sethi:2004np}\footnote{Some evidence that analogous statements should hold for higher-dimensional gauge theories was given in \cite{Chepelev:1997fk,Dine:1998am}.}. We extend these arguments to the \(1+1\)-dimensional $\mathcal{N}=(8,8)$ SYM theory needed for matrix string theory. The result implies that the $v^4/r^6$ term in the two-dimensional Coulomb-branch effective action is one-loop exact. We can therefore compute this term in the UV gauge theory, compare it with the IR orbifold description of the same amplitude, and fix the magnitude of the DVV Wilson coefficient. On the supergravity side this term appears both in the D1-brane probe action and in the Aichelburg--Sexl shockwave \cite{1971GReGr...2..303A} generated by light-cone momentum, making it natural to associate the interaction with graviton exchange.

The flow from the $\mathcal{N}=(8,8)$ SYM theory to the orbifold CFT in the IR defines an RG flow. A natural question is how to determine the Wilson coefficients of the leading irrelevant operators of the IR CFT. In the case at hand, both the UV and IR descriptions preserve maximal supersymmetry, so these deformations are highly constrained \cite{mstrevisited}. It is therefore plausible that the corresponding Wilson coefficients can be fixed precisely by computing protected quantities.

In section \ref{MST} we give a brief review of matrix string theory. In section \ref{dbi} we compute the $\mathcal{O}(v^4)$ interaction from gravity, first in the type-\Romannum{2}B background of extremal D1-branes and then in the type-\Romannum{2}A shockwave geometry. In section \ref{nonrenom}, we show that this interaction obeys a non-renormalization theorem on the gauge-theory side\footnote{We describe an extension of the argument presented in \cite{Paban:1998ea,Sethi:2004np}.}. In section \ref{sym} we perform the one-loop gauge-theory computation that fixes the coefficient of this interaction and show that it agrees with the gravity result, providing another check of the duality. Finally, in section \ref{symmorb} we review the four-graviton symmetric-orbifold amplitude \cite{Arutyunov:1997gi,Arutyunov:1998eq} generated by the DVV operator, extract its low-energy, large-impact-parameter limit, and fix the Wilson coefficient. An alternative check of this coefficient can also be found by matching to the expected string answer (see \cite{mstrevisited,principles}).

We end with a discussion of the result and some comments on how this technique can be used to fix the coefficients of other DVV-like deformation operators in the IR. The appendices collect details of the calculations and the various conventions used in the main text.

\section{Matrix String Theory}\label{MST}

The matrix string duality can be understood\footnote{For further details see \cite{lin2025tasilecturesmatrixtheory,principles}.} by considering the extremal $\rm{D}1$-brane solution in type-\Romannum{2}B:
\begin{align}\label{d1brane}
    &ds^2_{\rm{str}} =(\mathcal{H}(r))^{-\frac{1}{2}}\left(-dt^2+dx^2\right)+(\mathcal{H}(r))^{\frac{1}{2}}(dr^2+r^2d\Omega_{7}^2),\\
    & e^{\Phi} =  (\mathcal{H}(r))^{\frac{1}{2}},\qquad C_{2} = (\mathcal{H}(r))^{-1} dt \wedge dx\\
    & \mathcal{H}(r) = 1+\frac{32N_{1}\pi^2 g_{B}l_{B}^6}{r^6}
\end{align}
Here \(l_{B}\) is the type-\Romannum{2}B string length and \(g_{B}\) is the type-\Romannum{2}B string coupling. The holographic dictionary \cite{Itzhaki_1998} gives a dual description in terms of two-dimensional \(U(N)\), \(\mathcal N=(8,8)\) SYM:
\begin{align}
S_{\rm MSYM_{2}}
&=
\frac{1}{g_{\rm YM}^2}
\int d\tau d\sigma\,
\mathrm{Tr}\,\Big[
\frac14F_{\alpha\beta}F_{\alpha\beta}
+
\frac12D_\alpha X^iD_\alpha X^i
-
\frac14[X^i,X^j]^2
\\
&\hspace{2.2cm}
+\lambda^\alpha_+D_-\lambda^\alpha_+
+\lambda^{\dot\alpha}_-D_+\lambda^{\dot\alpha}_-
+\lambda^\alpha_+\gamma^i_{\alpha\dot\alpha}
[X^i,\lambda^{\dot\alpha}_-]
\Big],
\label{eq:MST-review-action}
\end{align}
where \(i=1,\ldots,8\), \(\alpha,\beta=0,1\), and
\(\alpha,\dot\alpha\) are chiral and anti-chiral \(SO(8)_R\)
spinor indices, and \(g_{\rm YM}^2=g_{B}/(2\pi l_{B}^2)\). In the IR regime the dilaton grows and the string coupling is large, so one applies S-duality to the background in \eqref{d1brane}, which gives the \(\rm F1\) background:
\begin{align*}
       &ds^2_{\rm{str}} =(\mathcal{H}(r))^{-1}\left(-dt^2+dx^2\right)+dr^2+r^2d\Omega_{7}^2,\\
    & e^{\Phi} =  (\mathcal{H}(r))^{-\frac{1}{2}},\qquad B_{2} = (\mathcal{H}(r))^{-1} dt \wedge dx\\
    & \mathcal{H}(r) = 1+\frac{32\pi^2 \tilde{g}_{B}^2\tilde{l}_{B}^6 N_{1}}{r^6}, \qquad \tilde{g}_{B} = 1/g_{B},\ \tilde{l}_{B} = g_{B}^{\frac{1}{2}}l_{B}
\end{align*}
Compactifying on a circle \(x\sim x+2\pi R\) and applying T-duality leads to type-\Romannum{2}A in the pp-wave spacetime
\begin{equation}
    ds_{\rm{str}}^2 =2dt\,d\tilde{x}+ \frac{32\pi^2g_{A}^2l_{A}^4R^2N_{1}}{r^6}d\tilde{x}^2+dr^2+r^2d\Omega_{7}^2
\label{ppwave}
\end{equation}
with
\begin{equation}
    \tilde{l}_{B} = l_{A}, \quad  g_{A} =  \tilde{g}_{B}\frac{\tilde{l}_{B}}{R}
\end{equation}
Thus the gauge-theory coupling is related to the type-\Romannum{2}A coupling by
\begin{equation*}
    g_{A} = \frac{1}{\sqrt{2\pi} g_{\rm YM}R}
\end{equation*}
In the strong-coupling limit the fields are projected onto the Cartan directions:
\begin{equation*}
[X^i,X^j]=0,
\qquad
[D_\sigma,X^i]=0,
\end{equation*}
so the surviving low-energy degrees of freedom may be simultaneously
diagonalized.  The off-diagonal modes become massive and decouple, leaving a
free theory of the \(N\) Cartan components together with their fermionic
partners and the decoupled abelian gauge multiplet.  Writing the commuting configuration in terms of the common transverse coordinate \(x^i\), defined in appendix \ref{transcoord},
\begin{equation*}
x^i
=
{\rm diag}\!\left(x^i_{(1)},\ldots,x^i_{(N)}\right),
\end{equation*}
the infrared action reduces to \(N\) free Green--Schwarz transverse fields on
the cylinder.  This diagonal form is not unique: after fixing to the Cartan
there remains the residual Weyl group \(S_N\subset U(N)\), which permutes the
eigenvalues.  The classical moduli space is therefore
\begin{equation*}
\mathcal M=\mathrm{Sym}^N(\mathbb R^8)
=
(\mathbb R^8)^N/S_N .
\end{equation*}
The full matrix fields \(X^i(\sigma)\) are single-valued on
the cylinder, but the diagonal entries themselves need only return to
themselves up to a Weyl permutation after \(\sigma\to \sigma+2\pi R\).  Thus
one may work in twisted sectors with boundary conditions
\begin{equation*}
x^i_{(a)}(\sigma+2\pi R)=x^i_{(g(a))}(\sigma),
\qquad
g\in S_N.
\end{equation*}
This is the residual gauge invariance in the Weyl group.  If the
permutation decomposes into cycles,
\([g]=\prod_{K=1}^N (K)^{n_K}\) with \(\sum_K K\,n_K=N\), then each cycle of
length \(K\) ties together \(K\) eigenvalues into a single field living on a
long cylinder of circumference \(2\pi K R\). The short cylinders associated with the individual
eigenvalues are sewn into longer cylinders by the Weyl twist.  A \(K\)-cycle
is therefore interpreted as one light-cone string carrying
\begin{equation*}
p_K^+=\frac{K}{R_-}=\frac{KR}{l_A^2}.
\end{equation*}
In the dimensionless orbifold convention used later, \(l_A^2=1\), this reduces to \(p_K^+=KR\).
The asymptotic free-string states are therefore labeled by conjugacy classes of
\(S_N\), or equivalently by collections of strings with \(n_K\) strings of
length \(K\).  The infrared fixed point is therefore the free symmetric-product orbifold
\begin{equation*}
U(N)\ {\rm SYM}
\xrightarrow[g_{\rm YM}R\to\infty]{}
\left(\mathbb R^8\right)^N/S_N,
\end{equation*}
with the supersymmetric \((R,R)\) spin structure inherited from the Yang--Mills
description.  In particular, there is no interacting realization of the
two-dimensional \(\mathcal N=(8,8)\) superconformal algebra at the fixed
point \cite{Seiberg_1998}, so the orbifold theory is genuinely free and different twisted sectors
are orthogonal in the Hilbert space. Moving away from this IR CFT along the RG flow can be viewed
as conformal perturbation theory around this orbifold fixed point,
\begin{equation}
\mathcal L_{\rm eff}
=
\mathcal L_{\rm orbifold}
\,+\,
\sum_a \lambda_a \mathcal O_a,
\label{eq:MST-review-conformal-perturbation}
\end{equation}
where the allowed \(\mathcal O_a\)\footnote{The operators \(\mathcal{O}_a\) have the CFT normalization.} must respect \(\mathcal N=(8,8)\)
supersymmetry and the transverse \(\mathfrak{so}(8)\) symmetry. Along this RG flow, the Wilson coefficients \(\lambda_{a}\) of the leading irrelevant operators of the IR CFT are not known a priori, but the operators \(\mathcal{O}_{a}\) are highly constrained. At leading order it can be proved \cite{Dijkgraaf:1997vv} that there is only one compatible vertex\footnote{More recent results also constrain further deformations \cite{mstrevisited}.}. This vertex is the analogue of the Mandelstam string vertex in LCSFT \cite{Mandelstam:1985ww,Dijkgraaf:2003nw}, which generates the superstring expansion in the light-cone gauge by explicitly
constructing the elementary joining/splitting interaction. The perturbative
matching between this orbifold description and spacetime string interactions (up to the undetermined coefficient)
was developed in \cite{Arutyunov:1997gi,Dijkgraaf:2003nw}. This leading vertex has the form
\begin{equation}
\Delta S
=
\frac{\lambda}{g_{\rm YM}}
\int d^2z\,\Sigma,
\qquad
\Sigma=\sum_{1\le I<J\le N}\Sigma_{(IJ)},
\label{eq:MST-review-DVV}
\end{equation}
where each \(\Sigma_{(IJ)}\) implements the simple transposition exchanging two
eigenvalues \(I\leftrightarrow J\).  Concretely, one isolates the relative
supermultiplet
\begin{equation*}
x_-^i=x^i_{(I)}-x^i_{(J)},
\qquad
\theta_-^a=\theta^a_{(I)}-\theta^a_{(J)},
\end{equation*}
on which the transposition acts as a \(\mathbb Z_2\) reflection.  The bosonic
twist field \(\sigma_{(IJ)}\) is characterized by
\begin{equation*}
\partial x_-^i(z)\,\sigma_{(IJ)}(0)
\sim
z^{-1/2}\,\tau^i_{(IJ)}(0),
\end{equation*}
while the fermionic twist is supplied by spin fields obeying
\begin{equation*}
\theta_-^a(z)\,\Sigma^i_{(IJ)}(0)
\sim
z^{-1/2}\,\gamma^i_{a\dot a}\,
\Sigma^{\dot a}_{(IJ)}(0).
\end{equation*}
Since \(h(\sigma)=h(\Sigma^i)=1/2\) and \(h(\tau^i)=1\), the unique
\(\mathfrak{so}(8)\)-invariant left-moving operator of minimal weight that
creates a simple cycle change is
\begin{equation*}
V_{(IJ)}(z)=\tau^i_{(IJ)}(z)\,\Sigma^i_{(IJ)}(z),
\qquad
h\bigl(V_{(IJ)}\bigr)=\frac32,
\end{equation*}
and similarly for the right-movers.  The full DVV insertion is therefore
\begin{equation*}
\Sigma_{(IJ)}(z,\bar z)
=
V_{(IJ)}(z)\,\widetilde V_{(IJ)}(\bar z)
=
\bigl(\tau^i\Sigma^i\bigr)_{(IJ)}
\bigl(\widetilde\tau^j\widetilde\Sigma^j\bigr)_{(IJ)},
\end{equation*}
which has weights \((3/2,3/2)\).  This form is fixed by
three requirements.  First, the operator must implement a simple transposition,
because elementary joining/splitting changes the number of cycles by one.
Second, it must be \(SO(8)\)-invariant and compatible with the
\(\mathcal N=(8,8)\) supercharges.  Third, it should have the smallest
possible scaling dimension so that it is the leading deformation away from the
free orbifold point.  Equivalently, \(V_{(IJ)}\) is the
\(G_{-1/2}\)-descendant of the chiral primary \(\sigma_{(IJ)}
\Sigma^{\dot a}_{(IJ)}\), which makes its supersymmetric origin manifest.
Combining the left- and right-moving pieces gives total dimension
\(\Delta=3\).  In the conformal-perturbation language of
\eqref{eq:MST-review-conformal-perturbation}, this means that the coupling of
the integrated DVV vertex has dimension \(-1\), so it is naturally suppressed
by the inverse mass of the off-diagonal \(W\)-bosons that have been integrated
out.  Since that scale is set by \(g_{\rm YM}\), the coefficient is of order
\(1/g_{\rm YM}\), equivalently one power of the type-IIA string coupling.
This is precisely the scaling expected for the elementary cubic
joining/splitting vertex, and the same twist-field insertion reproduces the full Mandelstam light-cone
interaction \cite{Dijkgraaf:2003nw}.  The rest of the paper fixes the
coefficient \(\lambda\) by matching the common long-distance \(v^4/r^6\)
interaction in the DBI, one-loop SYM, and orbifold
descriptions.

Although our main concern in this paper is the zero-flux sector relevant for
perturbative matrix strings, it is worth noting that the same \(1+1\)-dimensional
maximally supersymmetric gauge theory also admits a rich decomposition into
electric-flux sectors.  In the large-\(N\) matrix-string interpretation these
sectors are naturally related to D0-brane charge and its bound states
\cite{Witten:1995im,Kologlu:2016aev,Giddings_1999,cho2026fluxsectorsmatrixstring}.
Closely related two-dimensional SYM descriptions also appear in NCOS
decoupling limits with near-critical electric field
\cite{Seiberg:2000ms,Gopakumar:2000na,Klebanov:2000pp}.

\section{Gravity calculation}\label{dbi}

Starting from the background \eqref{d1brane}, we first perform a type-\Romannum{2}B calculation with \(N_{2}\) probe D1-branes moving in a background sourced by \(N_{1}\) D1-branes, taking \(N_{1}\gg N_{2}\) so that the probe back-reaction can be neglected. We then calculate the same quantity in the type-\Romannum{2}A pp-wave spacetime in \eqref{ppwave}. In this section the transverse coordinate is denoted by \(Y^i\). The SYM scalar in the ultraviolet gauge theory is denoted by \(X^i\), and later the orbifold scalar is denoted by \(x^i\). The normalizations are
\begin{equation}
Y^i=2\pi l_{B}^2X^i=l_{A}\,x^i .
\label{eq:coordinate-dictionary-calculations}
\end{equation}
 For more details on our conventions see appendix \ref{app:conventions}.

\subsection{\texorpdfstring{Type-\Romannum{2}B D1-D1 calculation}{Type-IIB D1-D1 calculation}}
In this section we consider \(N_{2}\) probe D1-branes moving in a background sourced by \(N_{1}\) D1-branes wrapped on a circle, \(z\sim z+2\pi R\), with \(N_{1}\gg N_{2}\).\footnote{One could also consider a multi-centered solution, which would be relevant for the scattering problem \cite{PhysRevLett.59.1617}.} The
string-frame D1 background is \cite{Itzhaki_1998}
\begin{align*}
&ds_{\rm str}^2= \left(1+\frac{32\pi^2g_BN_1 l_{B}^6}{Y^6}\right)^{-1/2}
\eta_{ab}\,d\sigma^a d\sigma^b
+
\left(1+\frac{32\pi^2g_BN_1l_{B}^6}{Y^6}\right)^{1/2}
dY^idY^i,
\\
&e^\Phi
=
\left(1+\frac{32\pi^2g_BN_1l_{B}^6}{Y^6}\right)^{1/2},\quad
C_{tz}
= \left[
\left(1+\frac{32\pi^2g_BN_1l_{B}^6}{Y^6}\right)^{-1}-1
\right].
\end{align*}
The \(N_2\)-D1 probe action is
\begin{equation}
S_{\rm D1}
=
-N_2\mu_1g_{B}^{-1}\int dt\,dz\,
 e^{-\phi}\sqrt{-\det P[G]}
+
N_2\mu_1 g_{B}^{-1}\int P[C_2],
\qquad
\mu_1=\frac{1}{2\pi l_{B}^2}.
\label{eq:D1-probe-action-full-factor}
\end{equation}
Let the transverse embedding be \(Y^i=Y^i(t,z)\).  The pullback metric is
\begin{equation*}
P[G]_{ab}
=
G_{ab}+G_{ij}\partial_aY^i\partial_bY^j.
\end{equation*}
The probe action is
\begin{align*}
S_{\rm D1}
&=
-N_2\mu_1g_B^{-1}\int dt\,dz\,
\mathcal H^{-1}
\sqrt{
-\det
\left[
\eta_{ab}
+
\mathcal H
\partial_aY^i\partial_bY^i
\right]
}
\nonumber\\
&\quad
+
N_2\mu_1g_B^{-1}\int dt\,dz\,
\left[
\mathcal H^{-1}-1
\right].
\label{eq:D1-full-probe-action-before-zero-mode}
\end{align*}
For the scattering zero mode,
\begin{equation*}
Y^i=Y^i(t),
\qquad
\partial_zY^i=0,
\qquad
\partial_tY^i\partial_tY^i=v_Y^2.
\end{equation*}
After integrating over the wrapped circle, \(\int dz=2\pi R\), the Lagrangian is 
(the static terms cancel except for the rest mass):
\begin{equation*}
L_{\rm D1}
=
-\frac{N_2R}{g_Bl_{B}^2}
+
\frac{N_2R}{2g_Bl_{B}^2}v_Y^2
+
\frac{N_2R}{8g_Bl_{B}^2}
\mathcal H(Y)v_Y^4
+O(v_Y^6).
\end{equation*}
The free \(N_1=0\) contribution is
\begin{equation}
L_{\rm free}
=
-\frac{N_2R}{g_Bl_{B}^2}
+
\frac{N_2R}{2g_Bl_{B}^2}v_Y^2
+
\frac{N_2R}{8g_Bl_{B}^2}v_Y^4
+O(v_Y^6).
\end{equation}
Thus the leading interaction is
\begin{equation}
\Delta L_{\rm D1}
=
L_{\rm D1}-L_{\rm free}
=
4\pi^2RN_1N_2l_{B}^4\frac{v_Y^4}{Y^6}
\end{equation}
Using \eqref{eq:appendix-alpha0-alpha-gym-from-mtheory} and \eqref{eq:coordinate-dictionary-calculations}, we obtain
\begin{equation}
\Delta L_{\rm D1}
=
2\pi R\frac{N_1N_2}{g_{\rm YM}^2}\frac{v^4}{r^6}.
\label{eq:DBI-common-potential}
\end{equation}

\subsection{\texorpdfstring{Type-\Romannum{2}A/Shockwave calculation}{Type-IIA/Shockwave calculation}}
We now do a type-\Romannum{2}A calculation in the pp-wave spacetime of section \ref{MST}. Another way to get this spacetime \cite{Becker_1998,Becker:1997xw} is to consider the background sourced by a graviton in ten dimensions; this is the
Aichelburg--Sexl shockwave geometry \cite{1971GReGr...2..303A,Dray:1984ha} sourced by light-cone momentum.  We use
\(x^\pm=x^9\pm t\) and choose the light-cone time
\(\tau=\frac12x^+\).  A source graviton carrying light-cone momentum \(p_1^+\),
localized at \(x^-=0\) and at the origin of the eight transverse directions,
produces
\begin{equation}
ds^2
=
dx^+dx^-+dY^idY^i+h_{--}(x^-,Y)(dx^-)^2,
\qquad
Y^2=Y^iY^i.
\end{equation}
The Green function in eight transverse dimensions is
\begin{equation*}
G_8(Y)=\frac{1}{6\omega_7Y^6}=\frac{1}{2\pi^4Y^6},
\qquad
\omega_7=\frac{\pi^4}{3}.
\end{equation*}
The linearized Einstein equation gives
\begin{equation*}
h_{--}(x^-,Y)
=
\frac{2\kappa_{10}^2p_1^+}{6\omega_7Y^6}\delta(x^-).
\end{equation*}
Averaging over the compact light-cone circle \(x^-\sim x^-+2\pi R_-\),
\begin{equation*}
h_{--}(Y)
=
\frac{\kappa_{10}^2p_1^+}{6\pi R_-\omega_7Y^6}
=
\frac{4G_{10}p_1^+}{\pi^4R_-Y^6},
\label{eq:shockwave-profile-our-conventions}
\end{equation*}
where \(\kappa_{10}^2=8\pi G_{10}\). When translated to the gauge-theory variables, this background is the same pp-wave spacetime used in section \ref{MST}. For a probe graviton in this background with fixed \(p_2^+\), we start from the massive particle
action
\begin{equation}
S
=
-m\int d\tau\,
\left(
-2\dot x^- - v_Y^2 - h_{--}\dot x^{-2}
\right)^{1/2},
\qquad
v_Y^2=\dot Y^i\dot Y^i.
\end{equation}
The momentum conjugate to \(x^-\) is
\begin{equation*}
p_2^+
=
m\,
\frac{1+h_{--}\dot x^-}
{\left(-2\dot x^- - v_Y^2 - h_{--}\dot x^{-2}\right)^{1/2}}.
\end{equation*}
At fixed \(p_2^+\) we take the null limit \(m\to0\), which imposes
\begin{equation}
-2\dot x^- - v_Y^2 - h_{--}\dot x^{-2}=0.
\end{equation}
Solving this equation for the branch that is smooth at \(h_{--}\to0\),
\begin{equation*}
\dot x^-
=
\frac{\sqrt{1-h_{--}v_Y^2}-1}{h_{--}}.
\end{equation*}
Taking the Legendre transform to get the fixed-momentum interaction, we have
\begin{equation}
L'(p_2^+)
=
-p_2^+\dot x^-.
\end{equation}
Using the expansion
\begin{equation*}
\frac{1-\sqrt{1-h_{--}v_Y^2}}{h_{--}}
=
\frac12v_Y^2+\frac18h_{--}v_Y^4+O(h_{--}^2v_Y^6),
\end{equation*}
we find the leading interaction is
\begin{equation}
\Delta L_{\rm shock}
=
\frac{p_2^+}{8}h_{--}v_Y^4
=
\frac{G_{10}p_1^+p_2^+}{2\pi^4R_-}\frac{v_Y^4}{Y^6}
=
2\pi R\frac{N_1N_2}{g_{\rm YM}^2}\frac{v^4}{r^6}
\label{eq:10d-probe-graviton}
\end{equation}
This reproduces the answer from the D1-D1 calculation, \eqref{eq:DBI-common-potential}. It also shows that the D1-D1 calculation can be interpreted as graviton exchange, which will be useful when we compare to the orbifold calculation.

\section{Non-renormalization theorem for one-loop}\label{nonrenom}
In this section we prove the non-renormalization theorem needed for the matrix-string matching. The argument is an extension of the non-renormalization theorems in \cite{Paban:1998ea,Sethi:2004np}, adapted to the \(1+1\)-dimensional theory appropriate to matrix string theory. In particular, the fermions transform as \(8_s\oplus 8_c\) under the transverse \(\mathrm{Spin}(8)\), and the eight-fermion top component has a different tensor structure from the \(0+1\)-dimensional \(\mathrm{Spin}(9)\) case. We show that the supersymmetric Ward identity fixes this top component uniquely and therefore forces the scalar four-derivative coefficient to be harmonic on the eight-dimensional transverse moduli space.

Consider a generic point on the Coulomb branch of the \(SU(2)\) theory, or more generally one pair of separated eigenvalue clusters. We refer to \(X^i\) as the ultraviolet SYM scalar, while \(x^i\) is the matrix-string-normalized scalar, related by
\[
X^i=\frac{g_{\rm YM}}{\sqrt{2\pi}}x^i.
\]
The relative Cartan fields form a free abelian \(\mathcal N=(8,8)\) multiplet with transverse bosons \(x^i\), \(i=1,\ldots,8\), and fermions \(\theta\). Away from coincident eigenvalues the off-diagonal fields have mass
\[
m_W\sim r_X=\frac{g_{\rm YM}}{\sqrt{2\pi}}r,
\qquad
r^2=x^ix^i,
\]
so the Wilsonian action for the light fields admits a derivative expansion. It is useful to organize the long-distance, low-velocity expansion as
\begin{equation}
\mathcal L_{\rm eff}=\mathcal L^{[0]}+\mathcal L^{[1]}+\mathcal L^{[2]}+\cdots,
\end{equation}
where
\[
    \mathcal{L} = c_{00}v^2+ \sum_{m,n=1}^{\infty}\frac{c_{mn}}{g_{\rm{YM}}^{2m}}\frac{v^{2n+2}}{r^{2m+4n}} .
\]
Here \(\mathcal L^{[\ell]}\) denotes the \(\ell\)-loop contribution; schematically \footnote{The $\mathcal{O}(v^2)$ term is also protected for sixteen supersymmetries.},
\begin{align}
\mathcal L^{[0]} &= c_{00}v^2,\nonumber\\
g_{\rm{YM}}^2\mathcal L^{[1]} &= \qquad c_{11}\frac{v^4}{r^6}
             + c_{12}\frac{v^6}{r^{10}}
             + c_{13}\frac{v^8}{r^{14}}+\cdots,\nonumber\\
g_{\rm{YM}}^4\mathcal L^{[2]} &= \qquad c_{21}\frac{v^4}{r^8}
             + c_{22}\frac{v^6}{r^{12}}
             + c_{23}\frac{v^8}{r^{16}}+\cdots .
\label{eq:velocity-large-r-expansion}
\end{align}
We show that the four-derivative coefficient is one-loop exact \footnote{The terms on the diagonal are exactly those that come from the supergravity expansion, it is also expected that $c_{12}$ =0 and the $\mathcal{O}(v^6)$ term is non-renormalized, to calculate the coefficient would require a two-loop computation \cite{Becker:1997wh}},
\begin{equation}
 c_{\ell 1}=0,\qquad \ell\ge 2,
\end{equation}
The fermions decompose under \(\mathrm{Spin}(1,1)\times \mathrm{Spin}(8)\) as \(\lambda_+^a\in 8_s\) and \(\lambda_-^{\dot a}\in 8_c\).
The leading scalar supersymmetry variation is
\begin{equation}
\delta x^i
=
i\epsilon_+^a\gamma^i_{a\dot a}\lambda_-^{\dot a}
+
i\epsilon_-^{\dot a}\gamma^i_{a\dot a}\lambda_+^a .
\label{eq:leading-scalar-susy-variation}
\end{equation}
We denote the four-derivative supersymmetric term by \(\mathcal L_2\), and label its components by fermion number:
\[
\mathcal L_2
=
\mathcal L_2^{(0)}
+\mathcal L_2^{(2)}
+\mathcal L_2^{(4)}
+\mathcal L_2^{(6)}
+\mathcal L_2^{(8)},
\qquad
\mathcal L_2^{(8)} \sim \lambda_+^4\lambda_-^4 .
\]
The full Lagrangian and supersymmetry transformations are expanded as
\[
\mathcal L=\mathcal L_1+\mathcal L_2+\cdots,
\qquad
\delta x^i=\delta_0x^i+\delta_2x^i+\cdots,
\qquad
\delta\lambda=\delta_0\lambda+\delta_3\lambda+\cdots .
\]
As in \cite{Paban:1998ea}, the corrected \(\delta x^i\) contains at most five fermions, while the corrected \(\delta\lambda\) contains at most six fermions.
Varying the eight-fermion top term gives
\begin{equation}
\delta\mathcal L_2^{(8)}
=
\delta x^i\,\partial_i\mathcal L_2^{(8)}
+
\delta\lambda\,\frac{\partial\mathcal L_2^{(8)}}{\partial\lambda}.
\end{equation}
The first term contains nine fermions,
\[
\delta_0x^i\,\partial_i\mathcal L_2^{(8)}\sim\epsilon\lambda^9,
\]
while the second contains seven,
\[
\delta_0\lambda\,\frac{\partial\mathcal L_2^{(8)}}{\partial\lambda}
\sim
\epsilon\,\partial x\,\lambda^7 .
\]
The seven-fermion terms can mix with variations of the six-fermion terms in \(\mathcal L_2\). The nine-fermion term cannot: no lower term in \(\mathcal L_2\) varies into nine fermions, and \(\delta\mathcal L_1\) cannot contain nine fermions because of the fermion-counting bounds above. Therefore the nine-fermion term must vanish by itself:
\begin{equation}
\delta_0x^i\,\partial_i\mathcal L_2^{(8)}=0 .
\label{eq:nine-fermion-constraint}
\end{equation}
Using the two parts of \(\delta_0x^i\), this gives the pair of first-order constraints
\begin{equation}
\gamma^i_{a\dot a}\lambda_-^{\dot a}\partial_i\mathcal L_2^{(8)}=0,
\qquad
\gamma^i_{a\dot a}\lambda_+^a\partial_i\mathcal L_2^{(8)}=0 .
\label{eq:two-first-order-constraints}
\end{equation}
\subsection{General Eight-Fermion term}
Define the \(SO(8)\) vector bilinear
\[
V^i=\lambda_+^a\gamma^i_{a\dot a}\lambda_-^{\dot a},
\qquad
\xi=x\cdot V=x^iV^i,
\qquad
\gamma=V^2=V^iV^i .
\]
The eight-fermion term belongs to \(\wedge^4 8_s\otimes \wedge^4 8_c\).
The relevant \(SO(8)\) Fierz decomposition leaves one scalar, one symmetric traceless rank-two tensor, and one symmetric traceless rank-four tensor. Since trace terms can be absorbed into radial functions of \(r\), the most general rotationally invariant top term may be written as
\begin{equation}
\mathcal L_2^{(8)}
=
a(r)\gamma^2+b(r)\gamma\xi^2+c(r)\xi^4 .
\label{eq:eight-fermion-ansatz}
\end{equation}
Start with \(\gamma^i_{a\dot a}\lambda_-^{\dot a}\partial_i\mathcal L_2^{(8)}=0\) and act with
\[
\gamma^j_{a\dot b}
\frac{\partial}{\partial\lambda_-^{\dot b}}
\partial_j .
\]
Using the \(SO(8)\) Clifford algebra and the fact that \(\mathcal L_2^{(8)}\) contains exactly four \(\lambda_-\)'s, one obtains
\begin{equation}
\Delta_{\mathbb R^8}\mathcal L_2^{(8)}=0 .
\label{eq:eight-fermion-laplace}
\end{equation}
Now apply the Laplacian to \(\mathcal L_2^{(8)}=a\gamma^2+b\gamma\xi^2+c\xi^4\). Since \(\partial_i\xi=V_i\) and \(\partial_i\gamma=0\), we have
\[
\Delta \xi^2=2\gamma,
\qquad
\Delta \xi^4=12\gamma\xi^2 .
\]
For a radial function in eight dimensions, \(\Delta h=h''+7h'/r\).
Therefore
\begin{equation}
\Delta\mathcal L_2^{(8)}
=
\left(a''+\frac7r a'+2b\right)\gamma^2
+
\left(b''+\frac{11}{r}b'+12c\right)\gamma\xi^2
+
\left(c''+\frac{15}{r}c'\right)\xi^4 .
\label{eq:eight-fermion-laplace-expanded}
\end{equation}
Independence of the three structures gives a set of coupled second-order equations. Solving these gives
\begin{align*}
&c(r)=C_0+\frac{C}{r^{14}}\\
&b(r)
=
B_0+\frac{B_1}{r^{10}}
-\frac{C_0}{2}r^2
-\frac{C}{2r^{12}}\\
&a(r)
=
A_0+\frac{A_1}{r^6}
-\frac{B_1}{8r^8}
+\frac{C}{40r^{10}}
-\frac{B_0}{8}r^2
+\frac{C_0}{40}r^4 
\end{align*}
Thus the Laplacian condition alone is weaker than full supersymmetry. Returning to the original nine-fermion equation \(\gamma^i_{a\dot a}\lambda_-^{\dot a}\partial_i\mathcal L_2^{(8)}=0\), define \(Q_a^i=\gamma^i_{a\dot a}\lambda_-^{\dot a}\) and choose
\[
x^i=(r,0,\ldots,0).
\]
The first-order constraint becomes
\begin{equation}
Q_a^1
\left[
a'\gamma^2
+
r^2b'\gamma(V^1)^2
+
r^4c'(V^1)^4
\right]
+
2rb\,\gamma V^1(Q_a\cdot V)
+
4r^3c\,(V^1)^3(Q_a\cdot V)
=0 .
\label{eq:first-order-constraint-component}
\end{equation}
The relevant \(SO(8)\) Fierz identity is (see Appendix \ref{Fierz} for the derivation):
\begin{equation}
Q_a^1\gamma^2
-24Q_a^1\gamma(V^1)^2
+56Q_a^1(V^1)^4
+4\gamma V^1(Q_a\cdot V)
-16(V^1)^3(Q_a\cdot V)
=0 .
\label{eq:so8-fierz-main}
\end{equation}
Therefore the coefficient vector must be proportional to the coefficient vector in this identity. Thus
\begin{equation}
r^2b'=-24a',
\qquad
r^4c'=56a',
\qquad
rb=2a',
\qquad
r^3c=-4a' .
\label{eq:first-order-radial-system}
\end{equation}
Solving this first-order system gives the unique decaying solution
\begin{equation}
a(r)=\frac{A}{r^{10}},
\qquad
b(r)=-\frac{20A}{r^{12}},
\qquad
c(r)=\frac{40A}{r^{14}} .
\label{eq:first-order-radial-solution}
\end{equation}
Therefore
\begin{equation}
\mathcal L_2^{(8)}
=
A\left[
\frac{(V^2)^2}{r^{10}}
-
20\frac{V^2(x\cdot V)^2}{r^{12}}
+
40\frac{(x\cdot V)^4}{r^{14}}
\right].
\label{eq:eight-fermion-final-form}
\end{equation}
Thus the top fermion term is fixed by supersymmetry. To relate it to the scalar coefficient, one needs more than the nine-fermion constraint alone: \(\mathcal L_2^{(8)}\) is the top component of the same four-derivative invariant whose bottom component is
\[
\mathcal L_2^{(0)}=f(x)(\partial x)^4 .
\]
The leading transformations have the schematic form \(\delta x^i\sim\epsilon\lambda^i\), \(\delta\lambda\sim \epsilon\,\partial x\). For example, varying the scalar dependence of the bottom term gives
\[
\delta\mathcal L_2^{(0)}
\supset
\partial_i f\,\delta x^i(\partial x)^4
\sim
(\partial_i f)(\epsilon\lambda^i)(\partial x)^4 .
\]
This one-fermion term must be cancelled by the variation of a two-fermion term. Since varying one fermion gives \(\delta\lambda\sim\epsilon\,\partial x\), the required two-fermion term has the schematic coefficient
\[
\mathcal L_2^{(2)}
\sim
\partial_i f\,\lambda^2(\partial x)^3 .
\]
Then \(\delta\lambda\) acting on \(\mathcal L_2^{(2)}\) produces
\[
\delta\mathcal L_2^{(2)}
\sim
(\partial_i f)\epsilon\lambda(\partial x)^4,
\]
which cancels the one-fermion variation of \(\mathcal L_2^{(0)}\). Repeating the same argument gives the descent chain
\[
\mathcal L_2^{(0)}\sim f(\partial x)^4,\qquad
\mathcal L_2^{(2)}\sim \partial f\,\lambda^2(\partial x)^3,\qquad
\mathcal L_2^{(4)}\sim \partial^2 f\,\lambda^4(\partial x)^2,
\]
\[
\mathcal L_2^{(6)}\sim \partial^3 f\,\lambda^6(\partial x),\qquad
\mathcal L_2^{(8)}\sim \partial^4 f\,\lambda^8 .
\]
In \(1+1\) dimensions Lorentz invariance requires the top term to contain four left-moving and four right-moving fermions. The natural \(SO(8)\) vector bilinear \(V^i=\lambda_+^a\gamma^i_{a\dot a}\lambda_-^{\dot a}\) contains one of each chirality, so a convenient representative of the top component is
\begin{equation}
\mathcal L_2^{(8)}
\propto
V^iV^jV^kV^l
\partial_i\partial_j\partial_k\partial_l f(r).
\label{eq:eight-fermion-descendant}
\end{equation}
For \(f(r)=r^{-p}\), differentiation gives
\begin{equation}
V^iV^jV^kV^l
\partial_i\partial_j\partial_k\partial_l r^{-p}
=
3p(p+2)\frac{\gamma^2}{r^{p+4}}
-
6p(p+2)(p+4)\frac{\gamma\xi^2}{r^{p+6}}
+
p(p+2)(p+4)(p+6)\frac{\xi^4}{r^{p+8}} .
\label{eq:radial-fourth-derivative}
\end{equation}
The unique top term in \eqref{eq:eight-fermion-final-form} has powers
\[
\frac{\gamma^2}{r^{10}},
\qquad
\frac{\gamma\xi^2}{r^{12}},
\qquad
\frac{\xi^4}{r^{14}} .
\]
Hence \(p+4=10\), \(p+6=12\), and \(p+8=14\), so all three structures give \(p=6\). Thus \(f(r)\propto 1/r^6\).
For \(p=6\), the relative coefficients are \(1:-20:40\),
which matches the top form above. Equivalently, the scalar coefficient is harmonic on the eight-dimensional transverse space: \(\Delta_{\mathbb R^8}f=0\). For radial \(f\), \(f''+7f'/r=0\).
Therefore
\[
f(r)=C_0+\frac{C}{r^6}.
\]
Discarding the constant term for a decaying interaction gives
\begin{equation}
f(r)=\frac{C}{r^6}.
\end{equation}
Thus, purely from the gauge theory Ward identity, the scalar term at four derivatives must scale as $1/r^6$ and receives no corrections beyond one loop.

\section{One-loop SYM}\label{sym}
In the previous section we saw that the coefficient of the \(v^4/r^6\) term receives no corrections beyond one loop. In this section we calculate this coefficient by a one-loop computation in the gauge theory, similar to \cite{Becker:1997wh,Becker:1997xw}\footnote{Since the coefficient is protected, the one-loop computation fixes it unambiguously.}.
We start with the Euclidean two-dimensional maximally supersymmetric Yang--Mills action
\begin{align}
S_{\rm SYM}
&=
\frac{1}{g_{\rm YM}^2}
\int d\tau d\sigma\,
{\rm Tr}\bigg[
\frac14F_{\alpha\beta}F_{\alpha\beta}
+\frac12D_\alpha X^iD_\alpha X^i
-\frac14[X^i,X^j]^2
\nonumber\\
&\hspace{3.0cm}
+\frac12\psi^T\Gamma^\alpha D_\alpha\psi
+\frac12\psi^T\Gamma^i[X^i,\psi]
\bigg],
\label{eq:sym-section-action}
\end{align}
where \(\alpha,\beta=\tau,\sigma\), \(i,j=1,\ldots,8\),
\(D_\alpha=\partial_\alpha+[A_\alpha,\cdot]\), and
\(F_{\alpha\beta}=\partial_\alpha A_\beta-\partial_\beta A_\alpha
+[A_\alpha,A_\beta]\).  In particular, the scalar kinetic term is normalized as
\((2g_{\rm YM}^2)^{-1}{\rm Tr}(D_\alpha X^iD_\alpha X^i)\).
We expand around a classical background
\begin{equation}
X^i=B^i+g_{\rm YM}Y^i,
\qquad
A_\alpha=g_{\rm YM}A_\alpha .
\label{eq:background-split}
\end{equation}
The center-of-mass components are free and will be ignored. For the scattering
process we take the background in the ultraviolet SYM scalar normalization.
The relative velocity and impact parameter in this normalization are denoted by
\(v_X\) and \(b_X\):
\begin{equation*}
B^1
=
\frac{i}{2}
\begin{pmatrix}
v_X\tau&0\\
0&-v_X\tau
\end{pmatrix},
\qquad
B^2
=
\frac{i}{2}
\begin{pmatrix}
b_X&0\\
0&-b_X
\end{pmatrix},
\label{eq:BB-background}
\end{equation*}
and take \(B^i\) to obey the classical equations of motion.  The
background-field gauge condition is
\begin{equation*}
\overline D_M A_M
=
\partial_\alpha A_\alpha+[B^i,Y^i]=0.
\label{eq:bg-gauge}
\end{equation*}
The gauge-fixing term is
\begin{equation*}
S_{\rm gf}
=
\int d\tau d\sigma\,
{\rm Tr}
\left(
\overline D_M A_M
\right)^2 .
\label{eq:gauge-fixing}
\end{equation*}
The full action is decomposed as
\begin{equation*}
S=S_Y+S_A+S_{\rm fermi}+S_{\rm ghost}.
\label{eq:S-decomp}
\end{equation*}
For the \(U(2)\) computation, we decompose the fields as follows:
\begin{align*}
&A_\alpha
=
\frac{i}{2}
\left(
A_{\alpha 0}{\bf 1}+A_{\alpha a}\sigma^a
\right)\\
&X^i
=
\frac{i}{2}
\left(
X^i_0{\bf 1}+X^i_a\sigma^a
\right)\\
&\psi
=
\frac{i}{2}
\left(
\psi_0{\bf 1}+\psi_a\sigma^a
\right)
\end{align*}
with $ a=1,2,3$. Define
\begin{equation}
r_X^2(\tau)
=
b_X^2+v_X^2\tau^2.
\label{eq:rX2-background}
\end{equation}
The quantity \(r_X\) is the instantaneous off-diagonal mass.
\subsection{Bosons}
The Euclidean quadratic action for the scalar fluctuations is
\begin{align}
S_Y^{(2)}
=
i\int d\tau d\sigma\,
{\rm Tr}\bigg[
&
\frac12
Y^i_1
\left(
\partial_\tau^2+\partial_\sigma^2-r_X^2
\right)
Y^i_1
+
\frac12
Y^i_2
\left(
\partial_\tau^2+\partial_\sigma^2-r_X^2
\right)
Y^i_2
\nonumber\\
&
+
\frac12
Y^i_3
\left(
\partial_\tau^2+\partial_\sigma^2
\right)
Y^i_3
\bigg]
+
S_{Y,{\rm int}} .
\label{eq:SY2}
\end{align}
Here \(i=1,\ldots,8\).  The cubic and quartic terms are
\begin{align}
S_{Y,{\rm int}}
=
i\int d\tau d\sigma\,{\rm Tr}\bigg[
&
- g_{\rm YM}
\epsilon_{a3x}\epsilon_{cbx}B^i_3
Y^j_aY^i_bY^j_c
\nonumber\\
&
-\frac{g_{\rm YM}^2}{4}
\epsilon_{abx}\epsilon_{cdx}
Y^i_aY^j_bY^i_cY^j_d
\bigg].
\label{eq:SYint}
\end{align}
The Euclidean quadratic action for the gauge fields is
\begin{align}
S_A^{(2)}
=
i\int d\tau d\sigma\,{\rm Tr}\bigg[
&
\frac12
A_{\alpha 1}
\left(
\partial_\tau^2+\partial_\sigma^2-r_X^2
\right)
A_{\alpha 1}
+
\frac12
A_{\alpha 2}
\left(
\partial_\tau^2+\partial_\sigma^2-r_X^2
\right)
A_{\alpha 2}
\nonumber\\
&
+
\frac12
A_{\alpha 3}
\left(
\partial_\tau^2+\partial_\sigma^2
\right)
A_{\alpha 3}
+
2\epsilon_{ab3}
\partial_\tau B^i_3
A_{0a}Y^i_b
\bigg]
+
S_{A,{\rm int}} .
\label{eq:SA2}
\end{align}
Since only \(B^1_3=v_X\tau\) is time-dependent,
\begin{equation}
\partial_\tau B^1_3=v_X,
\qquad
\partial_\tau B^i_3=0
\quad
(i\neq 1).
\end{equation}
Thus the mixing term is
\begin{equation}
2v_X\,\epsilon_{ab3}A_{0a}Y^1_b.
\label{eq:mixing-term}
\end{equation}
The interaction terms are
\begin{align}
S_{A,{\rm int}}
=
i\int d\tau d\sigma\,{\rm Tr}\bigg[
&
g_{\rm YM}\epsilon_{abc}
\partial_\alpha Y^i_aA_{\alpha b}Y^i_c
-
g_{\rm YM}
\epsilon_{a3x}\epsilon_{bcx}
B^i_3A_{\alpha a}A_{\alpha b}Y^i_c
\nonumber\\
&
-\frac{g_{\rm YM}^2}{2}
\epsilon_{abx}\epsilon_{cdx}
A_{\alpha a}Y^i_bA_{\alpha c}Y^i_d
\bigg].
\label{eq:SAint}
\end{align}
Diagonalizing the bosonic mass matrix in
\eqref{eq:SY2} and \eqref{eq:SA2}, we obtain:
\begin{enumerate}

\item Sixteen real bosons with
\begin{equation*}
m^2=r_X^2,
\end{equation*}
\item Two real bosons with
\begin{equation*}
m^2=r_X^2+2v_X,
\end{equation*}
\item Two real bosons with
\begin{equation*}
m^2=r_X^2-2v_X,
\end{equation*}
\item Ten massless bosons: the eight diagonal transverse scalar modes \(Y^i_3\)
and the two diagonal gauge-field modes \(A_{\tau3},A_{\sigma3}\). These are
Cartan modes and do not enter the off-diagonal interaction determinant.
\end{enumerate}
For the interacting off-diagonal determinant, the massless modes are dropped.

\subsection{Fermions}

Using the gamma-matrix decomposition
\begin{equation*}
\Gamma^0=\sigma_3\otimes {\bf 1}_{16\times16},
\qquad
\Gamma^i=i\sigma_1\otimes\gamma^i ,
\label{eq:gamma-decomp}
\end{equation*}
where the \(\gamma^i\) are real and symmetric.  Define
\begin{equation*}
\psi_+
=
\frac{1}{\sqrt2}
(\psi_1+i\psi_2),
\qquad
\psi_-
=
\frac{1}{\sqrt2}
(\psi_1-i\psi_2).
\label{eq:psipm}
\end{equation*}
The massive fermion action is
\begin{equation}
S_{\rm fermi}^{(2)}
=
i\int d\tau d\sigma\,
\psi_+^T
\left(
\partial_\tau+\Gamma^\sigma\partial_\sigma
-v_X\tau\,\gamma^1-b_X\,\gamma^2
\right)
\psi_- .
\label{eq:Sfermi2}
\end{equation}
Thus the fermion mass matrix is
\begin{equation*}
m_f
=
v_X\tau\,\gamma^1+b_X\,\gamma^2 .
\label{eq:mf}
\end{equation*}
Squaring the fermion operator gives
\begin{equation*}
-\partial_\tau^2-\partial_\sigma^2+r_X^2-v_X\gamma^1.
\end{equation*}
Diagonalizing \(\gamma^1\), one obtains eight real fermions with
\begin{equation*}
m^2=r_X^2+v_X,
\end{equation*}
and eight real fermions with
\begin{equation*}
m^2=r_X^2-v_X.
\end{equation*}
The diagonal fermions \(\psi_3\) are massless and do not contribute to
the off-diagonal one-loop potential.

\subsection{Ghosts}

The ghost action is
\begin{align}
S_{\rm ghost}
=
i\int d\tau d\sigma\bigg[
&
C_1^*
\left(
-\partial_\tau^2-\partial_\sigma^2+r_X^2
\right)
C_1
+
C_2^*
\left(
-\partial_\tau^2-\partial_\sigma^2+r_X^2
\right)
C_2
\nonumber\\
&
-
C_3^*
\left(
\partial_\tau^2+\partial_\sigma^2
\right)
C_3
+
S_{\rm ghost,int}
\bigg].
\label{eq:Sghost}
\end{align}
Hence there are two massive complex ghosts with
\begin{equation*}
m^2=r_X^2,
\end{equation*}
and one massless complex ghost, which is part of the diagonal sector
and is dropped from the off-diagonal determinant.

\subsection{Feynman rules}
The rules used below are background-field rules rather than ordinary
perturbation theory around the  vacuum. The classical profile
\(B^i\) is kept fixed, while the off-diagonal fluctuations are integrated out.
The terms quadratic in the off-diagonal fields are therefore not treated as
interactions: they define the Gaussian operator in the background. Thus the
propagators are the exact inverse propagators for fluctuations in this fixed
background, with mass \(r_X^2(\tau)=B^iB^i\) and with the velocity-dependent
mixing terms included by diagonalizing the quadratic action.  After the quadratic operator has been fixed, the one-loop contribution is a functional determinant. Equivalently, it can be expanded as a sum of one-particle-irreducible vacuum diagrams with no external lines. The local long-distance expansion is then an expansion of this determinant in slow derivatives of the background, or in the small velocity-dependent shifts of the
mass eigenvalues, not an expansion in external scattering vertices around
\(B^i=0\).
For the  long-distance expansion we only need the momentum-space
propagator
\begin{equation*}
\Delta_B(k\mid m^2)
=
\frac{1}{k^2+m^2},
\qquad
k^2=k_\tau^2+k_\sigma^2.
\label{eq:DeltaB-momentum}
\end{equation*}
The relevant values of \(m^2\) are \(r_X^2\), \(r_X^2\pm2v_X\), and
\(r_X^2\pm v_X\), according to the spectrum above.

The nonzero two-point functions for the massive scalar modes are
\begin{equation}
\left\langle
Y^i_aY^j_b
\right\rangle
=
\delta_{ab}\delta^{ij}
\Delta_B(k\mid r_X^2),
\qquad
a,b=1,2,
\qquad
i,j=2,\ldots,8.
\label{eq:YY-prop}
\end{equation}
The gauge-field and \(Y^1\) propagators are
\begin{equation}
\left\langle
A_{0a}A_{0b}
\right\rangle
=
\left\langle
Y^1_aY^1_b
\right\rangle
=
\frac12\delta_{ab}
\left[
\Delta_B(k\mid r_X^2+2v_X)
+
\Delta_B(k\mid r_X^2-2v_X)
\right],
\label{eq:AA-YY-mixed-prop}
\end{equation}
and
\begin{equation}
\left\langle
A_{01}Y^1_2
\right\rangle
=
-
\left\langle
A_{02}Y^1_1
\right\rangle
=
-\frac12
\left[
\Delta_B(k\mid r_X^2+2v_X)
-
\Delta_B(k\mid r_X^2-2v_X)
\right].
\label{eq:AY-prop}
\end{equation}
The ghost propagator is
\begin{equation}
\left\langle
C_aC_b^*
\right\rangle
=
\delta_{ab}
\Delta_B(k\mid r_X^2),
\qquad
a,b=1,2.
\label{eq:ghost-prop}
\end{equation}
The massive fermion propagator satisfies
\begin{equation}
\left(
-\partial_\tau+\Gamma^\sigma\partial_\sigma
+
v_X\tau\gamma^1+b_X\gamma^2
\right)
\Delta_F
=
\delta(\tau-\tau')\delta(\sigma-\sigma').
\end{equation}
This can be expressed through the bosonic
propagator:
\begin{equation}
\Delta_F
=
\left(
\partial_\tau+\Gamma^\sigma\partial_\sigma
+
v_X\tau\gamma^1+b_X\gamma^2
\right)
\Delta_B
\left(
r_X^2-v_X\gamma^1
\right).
\label{eq:fermion-prop}
\end{equation}
Thus the fermions contribute shifted propagators with \(r_X^2\pm v_X\).

\subsection{One-loop determinant}

The off-diagonal one-loop determinant is
\begin{equation}
Z_{\rm 1-loop}
=
\det{}^{-6}{\cal O}_0\,
\det{}^{-1}{\cal O}_{+2}\,
\det{}^{-1}{\cal O}_{-2}\,
\det{}^{4}{\cal O}_{+1}\,
\det{}^{4}{\cal O}_{-1},
\label{eq:Z1loop}
\end{equation}
where
\begin{equation*}
{\cal O}_a
=
-\partial_\tau^2-\partial_\sigma^2+r_X^2+a v_X.
\label{eq:Oa-def}
\end{equation*}
Therefore
\begin{align*}
\log Z_{\rm 1-loop}
=
&
-6\,{\rm Tr}\log{\cal O}_0
-
{\rm Tr}\log{\cal O}_{+2}
-
{\rm Tr}\log{\cal O}_{-2}
\nonumber\\
&
+
4\,{\rm Tr}\log{\cal O}_{+1}
+
4\,{\rm Tr}\log{\cal O}_{-1}.
\label{eq:logZ1loop}
\end{align*}
Writing ${\cal O}_a={\cal O}+a v_X,$
then
\begin{equation*}
{\rm Tr}\log({\cal O}+a v_X)
=
{\rm Tr}\log{\cal O}
+
{\rm Tr}\log(1+a v_X\,{\cal O}^{-1}).
\end{equation*}
Expanding,
\begin{equation*}
{\rm Tr}\log(1+a v_X\,{\cal O}^{-1})
=
\sum_{n=1}^{\infty}
\frac{(-1)^{n+1}}{n}
(a v_X)^n
{\rm Tr}\,{\cal O}^{-n}.
\label{eq:log-expansion-2}
\end{equation*}
Diagrammatically, \({\cal O}^{-1}\) is the propagator
\begin{equation*}
\Delta(k)
=
\frac{1}{k^2+r_X^2},
\end{equation*}
and \(a v_X\) is a zero-momentum mass-insertion vertex.  Thus the
\(n\)-insertion one-loop vacuum diagram for species \(a\) is
\begin{equation}
{\cal D}_{a,n}
=
c_a
\frac{(-1)^{n+1}}{n}
(a v_X)^n
\int\frac{d^2k}{(2\pi)^2}
\frac{1}{(k^2+r_X^2)^n}.
\label{eq:D-an}
\end{equation}
The determinant weights are
\begin{equation}
(a,c_a)
=
(0,-6),
\quad
(+2,-1),
\quad
(-2,-1),
\quad
(+1,4),
\quad
(-1,4).
\label{eq:weights}
\end{equation}
The unshifted exponent \(c_0=-6\) is the sum of the \(16\) real unshifted
bosons and the two complex massive ghosts, \(-16/2+2=-6\).  The shifted
exponents come from the two real bosons at \(r_X^2\pm2v_X\) and the fermions at
\(r_X^2\pm v_X\).
Therefore
\begin{equation}
{\cal D}_n
=
\frac{(-1)^{n+1}}{n}
v_X^n
\left(
\sum_a c_a a^n
\right)
I_n(r_X),
\label{eq:Dn}
\end{equation}
where
\begin{equation*}
I_n(r_X)
=
\int\frac{d^2k}{(2\pi)^2}
\frac{1}{(k^2+r_X^2)^n}.
\label{eq:In}
\end{equation*}
The lower terms cancel,
\begin{equation*}
\sum_a c_a a^n=0,
\qquad
n=0,1,2,3,
\end{equation*}
and the first nonzero term is \(n=4\):
\begin{equation}
\sum_a c_a a^4
=
(-1)(2)^4+(-1)(-2)^4+4(1)^4+4(-1)^4
=
-24.
\end{equation}
Hence
\begin{equation*}
{\cal D}_4
=
\frac{(-1)^5}{4}
v_X^4(-24)
I_4(r_X)
=
6v_X^4I_4(r_X).
\label{eq:D4}
\end{equation*}
The two-dimensional integral is
\begin{equation}
I_4(r_X)
=
\int\frac{d^2k}{(2\pi)^2}
\frac{1}{(k^2+r_X^2)^4}
=
\frac{1}{12\pi r_X^6}.
\end{equation}
Multiplying by \(N_1N_2\) off-diagonal sectors gives 
\begin{equation}
\Delta L_{\rm 1-loop}
=
\frac{2\pi RN_1N_2}{g_{\rm YM}^2}
\frac{v^4}{r^6}.
\label{eq:final-L-x}
\end{equation}

\section{Symmetric Orbifold description} \label{symmorb}
We now pass from the ultraviolet SYM scalar \(X^i\) used in the previous
section to the orbifold-normalized scalar \(x^i\).  The change of variables is
\begin{equation}
X^i=\frac{g_{\rm YM}}{\sqrt{2\pi}}x^i,
\qquad
\psi=\frac{g_{\rm YM}}{\sqrt{2\pi}}\theta .
\label{eq:orbifold-variable-change}
\end{equation}
In these variables the SYM action takes the form
\begin{align}
S_{\rm SYM}
&=
\int d\tau d\sigma\,{\rm Tr}\bigg[
\frac{1}{4g_{\rm YM}^2}F_{\alpha\beta}F_{\alpha\beta}
+\frac{1}{4\pi}D_\alpha x^iD_\alpha x^i
-\frac{g_{\rm YM}^2}{16\pi^2}[x^i,x^j]^2
\nonumber\\
&\hspace{3.0cm}
+\frac{1}{4\pi}\theta^T\Gamma^\alpha D_\alpha\theta
+\frac{g_{\rm YM}}{2(2\pi)^{3/2}}
\theta^T\Gamma^i[x^i,\theta]
\bigg].
\label{eq:orbifold-normalized-sym-action}
\end{align}
Thus in the \(g_{\rm YM}\to\infty\) limit the commutator terms become infinitely costly and force the fields onto the commuting Cartan branch. After diagonalizing the commuting matrices and quotienting by the residual Weyl group \(S_N\), the infrared
theory is the free symmetric product
\begin{equation}
S_{\rm orb}
=
\frac{1}{4\pi}
\int d\tau d\sigma\,
\sum_{a=1}^{N}
\left[
\partial_\alpha x^i_a\partial_\alpha x^i_a
+\theta_a^T\Gamma^\alpha\partial_\alpha\theta_a
\right],
\qquad
(x_a,\theta_a)\sim (x_{\pi(a)},\theta_{\pi(a)}),
\quad
\pi\in S_N .
\label{eq:free-orbifold-action}
\end{equation}
The interacting matrix-string theory is obtained by perturbing this fixed
point by the leading supersymmetric twist operator.
We perturb the symmetric-orbifold theory by the DVV operator
\begin{equation}
\Delta S
=
\frac{\lambda}{g_{\rm YM}}
\int d^2z\,\Sigma,
\end{equation}
where
\begin{equation}
\Sigma=\sum_{I<J}\Sigma_{(IJ)},
\qquad
\Sigma_{(IJ)}
=
\bigl(\tau^i\Sigma^i\bigr)_{(IJ)}
\bigl(\widetilde\tau^j\widetilde\Sigma^j\bigr)_{(IJ)}.
\end{equation}
Here \(\sigma_{(IJ)}\) is the \(\mathbb Z_2\) twist for the relative bosons,
\(\tau^i_{(IJ)}\) is its excited twist descendant defined by
\(\partial x_-^i\,\sigma_{(IJ)}\sim z^{-1/2}\tau^i_{(IJ)}\), and
\(\Sigma^i_{(IJ)}\), \(\Sigma^{\dot a}_{(IJ)}\) are the corresponding spin
fields for the relative fermions.  The left-moving factor
\(\tau^i\Sigma^i\) has weight \(3/2\), the right-moving factor has weight
\(3/2\), and the full operator is the unique \(\mathfrak{so}(8)\)-invariant
simple-transposition vertex of minimal dimension.  Equivalently, it is the
super-descendant of the chiral primary
\(\sigma_{(IJ)}\Sigma^{\dot a}_{(IJ)}\), which is why this particular
combination preserves the \(\mathcal N=(8,8)\) supersymmetry while coupling
different twisted sectors. For a connected four-point amplitude, the second-order perturbative
contribution is (here $\sigma \sim \sigma+2\pi$):
\begin{equation}
S_{\rm MST}^{(2)}
=
-\frac12
\left(
\frac{\lambda}{g_{\rm YM }R}
\right)^2
\int d^2z_1\,d^2z_2\,
|z_1||z_2|\,
\langle f|
T\{\Sigma(z_1)\Sigma(z_2)\}
|i\rangle.
\end{equation}
We use \(d^2z=2d\tau d\sigma\).
The orbifold computation is intrinsically written in dimensionless
nonrelativistic variables. The Mandelstam variables are
\begin{equation}
s=-(k_1+k_2)^2,
\qquad
t=-(k_2+k_3)^2,
\qquad
u=-(k_1+k_3)^2,
\qquad
s+t+u=0.
\label{eq:orbifold-momenta}
\end{equation}
The four-graviton amplitude computed in \cite{Arutyunov:1997gi,Arutyunov:1998eq}, in our notation then becomes:
\begin{equation}
\mathcal A_{\rm MST}^{(4)}
=
-\frac{\pi^3\lambda^2}{4g_{\rm YM}^2R^2}
K_{\rm NS}
\frac{
\Gamma\!\left(-\frac{s}{4}\right)
\Gamma\!\left(-\frac{t}{4}\right)
\Gamma\!\left(-\frac{u}{4}\right)
}{
\Gamma\!\left(1+\frac{s}{4}\right)
\Gamma\!\left(1+\frac{t}{4}\right)
\Gamma\!\left(1+\frac{u}{4}\right)
}
\label{eq:orbifold-amplitude}
\end{equation}
\subsection{Low energy, large impact parameter limit}

The reduced orbifold amplitude in \eqref{eq:orbifold-amplitude} is written
entirely in dimensionless nonrelativistic variables.  We therefore take the
fixed \(s\), low-\(t\) limit of the amplitude \eqref{eq:orbifold-amplitude},
with \(u=-s-t\).  Defining
\begin{equation}
x=\frac{s}{4},
\qquad
y=\frac{t}{4},
\end{equation}
the gamma-function factor in the amplitude is
\begin{equation}
F(s,t,u)
=
\frac{
\Gamma(-x)\Gamma(-y)\Gamma(x+y)
}{
\Gamma(1+x)\Gamma(1+y)\Gamma(1-x-y)
}.
\end{equation}
At fixed \(x\) and small \(y\),
\begin{equation}
F(s,t,u)
=
\frac{64}{s^2t}
+O(t).
\end{equation}
For the transverse polarization choice
\begin{equation}
\zeta_1=\zeta_2=p,
\qquad
\zeta_3=\zeta_4=q,
\qquad
p^2=q^2=0,
\qquad
p\cdot q=1,
\qquad
p\cdot k_i=q\cdot k_i=0,
\end{equation}
we find \(K_{\rm NS}=s^4/16\).  Hence
\begin{equation}\label{orbifoldfinal}
\mathcal A_{\rm MST}^{(4)}
=
-\frac{\pi^3\lambda^2}{g_{\rm YM}^2R^2}
\frac{s^2}{t}
\end{equation}
To compare this pole with the long-distance gauge-theory interaction we use the
Born relation in the light-cone normalization of
\cite{Becker_1998,lin2025tasilecturesmatrixtheory}.  For
external states normalized with \(2p^+\), and after averaging over the compact
light-cone circle of length \(2\pi R_-\), the momentum-space potential is
\begin{equation}
\widetilde V(q)
=
\frac{1}{2\pi R_-}\frac{1}{4p_1^+p_2^+}\,
\mathcal A_{\rm pole}(q).
\label{eq:born-rule-lightcone}
\end{equation}
This follows from
\begin{equation*}
    V(r) = \frac{1}{2\pi R_{-}}\prod_{i}\frac{1}{\sqrt{2E_{i}}}\int \frac{d^8q}{(2\pi)^8}\mathcal{A}(q)e^{i\vec{q}\cdot \vec{r}}
\end{equation*}
Here \(q\) is conjugate to the transverse orbifold coordinate \(r\), so
\(t=-q^2\).  Applying \eqref{eq:born-rule-lightcone} to
\eqref{orbifoldfinal} gives
\begin{equation}
\widetilde V_{\rm MST}(q)
=
\frac{1}{2\pi R_-}\frac{1}{4p_1^+p_2^+}
\mathcal A_{\rm MST,pole}^{(4)}
=
\frac{\pi^2\lambda^2}{8g_{\rm YM}^2R^2R_-\,p_1^+p_2^+}
\frac{s^2}{q^2},
\end{equation}
Using the eight-dimensional Fourier transform in the same orbifold coordinate,
\begin{equation}
\int d^8r\,e^{-iq\cdot r}
\frac{1}{r^6}
=
\frac{2\pi^4}{q^2},
\end{equation}
the corresponding coordinate-space potential is
\begin{equation}
V_{\rm MST}(r)
=
\frac{\lambda^2}{16\pi^2g_{\rm YM}^2R^2R_-\,p_1^+p_2^+}
\frac{s^2}{r^6}.
\end{equation}
Finally, in the dimensionless orbifold convention used in this section we set
\(l_{A}^2=1\), so \(R_-=l_{A}^2/R=1/R\).  The light-cone momentum and
nonrelativistic invariant are therefore
\begin{equation}
s=(p_1^+p_2^+)v^2,
\qquad
p_a^+=\frac{N_a}{R_-}=N_{a}R,
\qquad
R_-=\frac{1}{R},
\end{equation}
this becomes
\begin{equation}
V_{\rm MST}(r)
=
\frac{\lambda^2}{16\pi^2g_{\rm YM}^2}
R\,N_1N_2
\frac{v^4}{r^6}.
\label{eq:orbifold-potential}
\end{equation}

\subsection{Coefficient of DVV}
Matching the integrated zero-mode interaction
 to \eqref{eq:orbifold-potential} gives
\begin{equation}
\frac{\lambda^2}{16\pi^2}R=2\pi R,
\end{equation}
and therefore fixes the magnitude of the DVV coupling:
\begin{equation}\label{DVVcoefficient}
    \lambda = \pm 2^{5/2}\pi^{3/2}.
\end{equation}

\section{Discussion}\label{discussion}
In this paper we fixed the normalization of the leading matrix-string interaction by
matching a protected long-distance observable across the ultraviolet gauge theory, the
spacetime gravity description, and the infrared symmetric-orbifold description. The
observable used in the matching is the two-body large-impact-parameter interaction
\[
        \Delta L
        =
        \frac{2\pi RN_1N_2}{g_{\rm YM}^2}\frac{v^4}{r^6}.
\]
The same coefficient appears in the D1 probe calculation, the type-\Romannum{2}A shockwave computation, and the
one-loop effective action of the two-dimensional \(\mathcal{N}=(8,8)\), \(U(N)\) SYM theory. The non-renormalization argument then allows this UV one-loop result to
be extrapolated to the IR fixed point. Matching this protected term to the
second-order conformal perturbation by the DVV operator gives\footnote{We also note that this result is \(N\)-independent, while the matrix-string conjecture states that string theory is recovered in the large-\(N\) limit. The DVV operator is irrelevant in the RG flow at large \(N\), but at this order it is also the unique deformation at finite \(N\). This need not remain true for higher deformations.}
\[
        \lambda^2 = 32\pi^3,
        \qquad
        |\lambda| = 2^{5/2}\pi^{3/2}.
\]
Thus the magnitude of the elementary joining/splitting vertex in the symmetric-orbifold
description is fixed by the UV gauge theory. Since the four-point amplitude depends on two DVV insertions, the calculation fixes \(\lambda^2\), and therefore the magnitude of \(\lambda\), but not its sign. It would be interesting to fix the sign as well, perhaps by matching a UV observable closer to a three-point amplitude \cite{principles,mstrevisited}. The picture of string interactions from the UV gauge theory may be useful for this purpose \cite{Bonelli:1998yt,Wynter_2000,Giddings_1999}.

The usefulness of the calculation is that it gives a controlled way of extracting IR CFT data from the UV gauge theory. In a generic RG flow, matching an irrelevant deformation in the infrared to a short-distance gauge-theory calculation would not be reliable. One can try to apply the same strategy to other irrelevant twist-field deformations \cite{Dijkgraaf:2003nw,mstrevisited} in the symmetric-orbifold theory. These deformations may be related to other protected quantities in the gauge theory. For example, the \(\mathcal{O}(v^6)\) term is also expected to be protected\footnote{Some higher-derivative protected terms from the gauge-theory side may be relevant \cite{Paban:1998qy}.}. It may also be useful to study multi-body interactions in order to fix higher-twist interactions from the orbifold. If this expectation is correct, then one might be able to fix all such Wilson coefficients, which is an interesting prospect. We hope to report more on this in the future. Finally, it would be interesting to understand how the analysis is modified in nontrivial electric-flux sectors, where the same gauge theory describes states carrying additional D0-brane charge. Some processes with M-momentum transfer can also be studied perturbatively from the gauge theory \cite{Keski_Vakkuri_1998}; this provides a further non-perturbative check of the duality and a method for studying processes beyond perturbative string theory.\footnote{This approach was also used in \cite{Giddings:1998yd}. A particularly interesting process to study would be the one discussed in \cite{Sen:2025xaj}.}
\acknowledgments
I would like to specially thank Henry Lin for helpful discussions and valuable advice on the manuscript. I  would also like to thank Shiraz Minwalla, Chintan Patel, and  Herman Verlinde for the  very  useful discussions. I would also like to thank Minjae Cho, Chintan Patel, Savdeep Sethi and  Stephen Shenker for their comments on a preliminary version of this manuscript.

\appendix

\section{Conventions and notation}\label{app:conventions}
This appendix collects the notation used in the calculation. We use three transverse variables in different descriptions. The
physical spacetime coordinate is denoted by \(Y^i\), the ultraviolet SYM scalar
is denoted by \(X^i\), and the matrix-string or orbifold-normalized scalar is
denoted by \(x^i\).  The symmetric-orbifold eigenvalue fields are denoted by
\(x^i_{(a)}\) when it is useful to emphasize the permutation action of
\(S_N\).

\begin{table}[b]
\centering
\renewcommand{\arraystretch}{1.25}
\begin{tabular}{@{}L{0.15\textwidth}L{0.22\textwidth}L{0.27\textwidth}L{0.20\textwidth}@{}}
\hline
Description & Variables & Long-distance observable & Final orbifold form \\
\hline
DBI/D1 frame &
\(Y^i=2\pi l_{B}^2X^i\), \(R\), \(l_{B}\) &
zero-mode interaction from the probe DBI expansion &
\(V_0\) \\
Shockwave gravity &
\(Y^i\), \(p_a^+=N_a/R_-\), \(R_-=l_{A}^2/R\) &
probe-graviton Routhian in the Aichelburg--Sexl background &
\(V_0\) \\
One-loop SYM &
\(X^i=(g_{\rm YM}/\sqrt{2\pi})x^i\) &
protected one-loop four-derivative term after zero-mode integration &
\(V_0\) \\
Symmetric orbifold &
\(x^i\), \(q\) conjugate to \(r\), \(l_{A}^2=1\) &
DVV second-order pole converted to a potential by the Born rule &
\((\lambda^2/32\pi^3)V_0\) \\
\hline
\end{tabular}
\caption{Dictionary for the common long-distance observable used in the matching, with \(V_0=2\pi R\,N_1N_2g_{\rm YM}^{-2}v^4/r^6\).  The variable \(r\) and velocity \(v\) are always the orbifold-normalized separation and relative velocity in the final column.}
\label{tab:observable-dictionary}
\end{table}

\subsection{M-theory origin and the 9--11 flip}
In this appendix we describe matrix string theory starting from M-theory, following \cite{Motl:1997th,Seiberg:1997ad,Dijkgraaf:1997vv}.
Let \(\ell_p\) denote the
eleven-dimensional Planck length.  Before the 9--11 flip, the matrix model is
viewed as the DLCQ description of M-theory with light-cone radius \(R_-\), and
we also compactify one transverse spatial direction, denoted here by \(R_9\).
Reducing along the light-cone/M-theory circle gives the type-IIA D0-brane
frame; T-dualizing along the spatial circle gives the type-IIB D1-brane frame
used in the DBI calculation.  In this intermediate D1 frame,
\begin{equation}
l_{B}^2=\frac{\ell_p^3}{R_-},
\qquad
R=\frac{l_{B}^2}{R_9},
\qquad
g_B=\frac{R_-}{R_9}.
\label{eq:appendix-mtheory-to-D1}
\end{equation}
Here \(R\) is the radius of the D1 worldsheet circle on which the two-dimensional
gauge theory lives. The 9--11 flip exchanges which compact direction is regarded as the M-theory
circle.  After the flip, reduction on the circle of radius \(R_9\) gives the
perturbative type-IIA string whose slope we denote by \(l_{A}^2\):
\begin{equation}
l_{A}^2=\frac{\ell_p^3}{R_9}.
\label{eq:appendix-alpha-prime-from-mtheory}
\end{equation}
Combining \eqref{eq:appendix-mtheory-to-D1} and
\eqref{eq:appendix-alpha-prime-from-mtheory} gives the matrix-string light-cone
radius relation
\begin{equation}
l_{A}^2=l_{B}^2\frac{R_-}{R_9}=R R_-,
\qquad\Longleftrightarrow\qquad
R_-=\frac{l_{A}^2}{R}.
\label{eq:appendix-nine-eleven-radius}
\end{equation}
Thus the same integer \(N\) that counted D0-branes or D1 winding in the duality
chain becomes the discrete light-cone momentum quantum number,
\begin{equation}
p^+=\frac{N}{R_-}.
\label{eq:appendix-pplus-mtheory}
\end{equation}
This is the origin of the light-cone and radius conventions used in the
orbifold and shockwave calculations. Finally, the D1 worldvolume coupling is
\begin{equation}
g_{\rm YM}^2=\frac{g_B}{2\pi l_{B}^2},
\qquad
l_{A}^2 = 2\pi g_{\rm YM}^2l_{B}^4
\label{eq:appendix-gym-from-mtheory}
\end{equation}
Equivalently,
\begin{equation}
\frac{l_{B}^4}{l_{A}^2}=\frac{1}{2\pi g_{\rm YM}^2}.
\label{eq:appendix-alpha0-alpha-gym-from-mtheory}
\end{equation}
This relation is useful to convert between the D1-frame DBI normalization and the
matrix-string normalization.

\subsection{Transverse-coordinate normalizations}\label{transcoord}
The transverse-coordinate dictionary is
\begin{equation}
Y^i
=
2\pi l_{B}^2 X^i
=
l_{A}\,x^i .
\label{eq:appendix-coordinate-dictionary-start}
\end{equation}
The first equality is the standard D-brane normalization of a transverse scalar.  The
second equality defines the orbifold-normalized field \(x^i\), which is the
field whose eigenvalues become the transverse coordinates of the infrared
symmetric-orbifold theory.

Equation \eqref{eq:appendix-coordinate-dictionary-start} gives
\begin{equation}
X^i
=
\frac{l_{A}}{2\pi l_{B}^2}\,x^i .
\label{eq:appendix-X-pre-gym}
\end{equation}
Using the M-theory/D1/gauge-theory dictionary in
\eqref{eq:appendix-alpha0-alpha-gym-from-mtheory}, the ultraviolet SYM scalar and the orbifold scalar are related by
\begin{equation}
X^i
=
\frac{g_{\rm YM}}{\sqrt{2\pi}}x^i .
\label{eq:appendix-coordinate-dictionary}
\end{equation}
For two-body kinematics define
\begin{equation}
r_X^2=\Delta X^i\Delta X^i,
\qquad
v_X^2=\Delta\dot X^i\Delta\dot X^i,
\qquad
r^2=\Delta x^i\Delta x^i,
\qquad
v^2=\Delta\dot x^i\Delta\dot x^i,
\end{equation}
where the dot denotes differentiation with respect to the same worldsheet time.
Since the field redefinitions above are constant rescalings, the same factors
relate separations and velocities:
\begin{equation}
r_X
=
\frac{g_{\rm YM}}{\sqrt{2\pi}}r,
\qquad
v_X
=
\frac{g_{\rm YM}}{\sqrt{2\pi}}v,
\qquad
Y
=
\sqrt{l_{A}^2}\,r,
\qquad
v_Y
=
\sqrt{l_{A}^2}\,v .
\label{eq:appendix-rv-dictionary}
\end{equation}
Consequently,
\begin{equation}
\frac{v_X^4}{r_X^6}
=
\frac{2\pi}{g_{\rm YM}^2}\frac{v^4}{r^6},
\qquad
\frac{v_Y^4}{Y^6}
=
\frac{1}{l_{A}^2}\frac{v^4}{r^6}.
\label{eq:appendix-ratio-dictionary}
\end{equation}
These relations are the ones used when comparing the DBI, shockwave, SYM, and
orbifold answers.  In particular, the off-diagonal mass in the one-loop SYM
calculation is
\begin{equation}
r_X=\frac{g_{\rm YM}}{\sqrt{2\pi}}r.
\label{eq:appendix-mass-dictionary}
\end{equation}

\subsection{Radii, couplings, and light-cone momentum}
The spatial matrix-string circle has radius \(R\).  As derived from the
9--11 flip in \eqref{eq:appendix-nine-eleven-radius},
\begin{equation}
\sigma\sim\sigma+2\pi R,
\qquad
R_-=\frac{l_{A}^2}{R},
\qquad
p_a^+=\frac{N_a}{R_-}.
\label{eq:appendix-lightcone-dictionary}
\end{equation}
A cycle of length \(K\) in the symmetric product therefore carries
\begin{equation}
p_K^+=\frac{K}{R_-}=\frac{KR}{l_A^2}.
\label{eq:appendix-cycle-momentum}
\end{equation}
The coupling dictionary used in the text is
\begin{equation}
g_A=\frac{1}{\sqrt{2\pi}\,g_{\rm YM}R},
\qquad
G_{10}=8\pi^6 g_A^2 l_{A}^8
= \frac{4\pi^5l_{A}^8}{g_{\rm YM}^2R^2}.
\label{eq:appendix-coupling-dictionary}
\end{equation}
In the D1 frame we use the same relations summarized above,
\begin{equation}
l_{B}^2=\frac{\ell_p^3}{R_-},
\qquad
R=\frac{l_{B}^2}{R_9},
\qquad
g_B=\frac{R_-}{R_9},
\qquad
g_{\rm YM}^2=\frac{g_B}{2\pi l_{B}^2}.
\label{eq:appendix-D1-frame-dictionary}
\end{equation}
There are two different string lengths in these formulae because the calculation uses two dual descriptions separated by the 9--11 flip.  The parameter
\(l_{B}^2=\ell_p^3/R_-\) is the string slope in the intermediate type-IIB
D1-brane frame, and it is the slope that appears in the D1 DBI action and in
the physical D-brane scalar normalization \(Y^i=2\pi l_{B}^2X^i\).  The
parameter \(l_{A}^2=\ell_p^3/R_9\) is the final
type-IIA/matrix-string slope after the 9--11 flip; it sets the matrix-string
light-cone radius through \(R_- = l_{A}^2/R\) and the IR normalization
\(Y^i=l_{A}x^i\).  They should therefore not be identified before
using the duality dictionary.  The relation
\begin{equation}
\frac{l_{B}^4}{l_{A}^2}=\frac{1}{2\pi g_{\rm YM}^2}
\label{eq:appendix-D1-dictionary}
\end{equation}
is just \eqref{eq:appendix-alpha0-alpha-gym-from-mtheory}.

\section{\texorpdfstring{Derivation of the \(SO(8)\) Fierz Identity}{Derivation of the SO(8) Fierz Identity}}\label{Fierz}

We need the Fierz relation among the five nine-fermion structures
\begin{align}
S_1&=Q_a^1\gamma^2,
&
S_2&=Q_a^1\gamma(V^1)^2,
&
S_3&=Q_a^1(V^1)^4,
\nonumber\\
S_4&=\gamma V^1(Q_a\cdot V),
&
S_5&=(V^1)^3(Q_a\cdot V),
\label{eq:app-fierz-structures}
\end{align}
where \(Q_a^i=\gamma^i_{a\dot a}\lambda_-^{\dot a}\),
\(V^i=\lambda_+^b\gamma^i_{b\dot b}\lambda_-^{\dot b}\), and
\(\gamma=V^iV^i\). We work at fixed spinor index \(a=1\). The final
identity is \(SO(8)\)-covariant, so fixing \(a=1\) is only a convenient
way to determine the relative coefficients.
\subsection{Gamma-Matrix Convention}
We use \(16\times16\) Euclidean \(SO(8)\) gamma matrices

\begin{equation}
\Gamma^{2k-1}
=
\sigma_3^{\otimes(k-1)}
\otimes\sigma_1
\otimes \mathbf 1^{\otimes(4-k)},
\qquad
\Gamma^{2k}
=
\sigma_3^{\otimes(k-1)}
\otimes\sigma_2
\otimes \mathbf 1^{\otimes(4-k)},
\qquad
k=1,\ldots,4,
\label{eq:app-gamma-convention}
\end{equation}
with the standard Pauli matrices \(\sigma_i\). The chirality matrix is
\(\Gamma^9=\Gamma^1\Gamma^2\cdots\Gamma^8\). Choosing bases for the
\(+1\) and \(-1\) eigenspaces of \(\Gamma^9\), the gamma matrices take
the chiral form

\begin{equation}
\Gamma^i
=
\begin{pmatrix}
0 & \gamma^i_{a\dot a}\\
(\gamma^i)^T_{\dot a a} & 0
\end{pmatrix}.
\label{eq:app-chiral-gamma-block}
\end{equation}

The \(8\times8\) blocks \(\gamma^i_{a\dot a}\) obey

\begin{equation}
\gamma^i_{a\dot a}\gamma^j_{a\dot b}
+
\gamma^j_{a\dot a}\gamma^i_{a\dot b}
=
2\delta^{ij}\delta_{\dot a\dot b}.
\label{eq:app-so8-clifford}
\end{equation}
We denote the Grassmann variables by \(\eta_a=\lambda_+^a\) and
\(\chi_{\dot a}=\lambda_-^{\dot a}\), and use the canonical ordering
\(\eta_1<\eta_2<\cdots<\eta_8<\chi_1<\chi_2<\cdots<\chi_8\). All signs
below come from reordering products of Grassmann variables into this
canonical order. In this chiral basis, after a relabeling of the
\(8_s\) and \(8_c\) bases, one has \(Q_1^1=\chi_5\) and

\[
V^1
=
\eta_1\chi_5+\eta_2\chi_6+\eta_3\chi_7+\eta_4\chi_8
+\eta_5\chi_1+\eta_6\chi_2+\eta_7\chi_3+\eta_8\chi_4.
\]
The parts of \(\gamma=V^iV^i\) needed for the coefficient extraction are
\begin{align*}
\gamma\supset{}&
8\eta_1\eta_2\chi_1\chi_2
+8\eta_1\eta_3\chi_1\chi_3
+8\eta_1\eta_4\chi_2\chi_3\\
&+8\eta_2\eta_3\chi_1\chi_4
+8\eta_2\eta_4\chi_2\chi_4
+8\eta_3\eta_4\chi_3\chi_4\\
&+4\eta_2\eta_7\chi_4\chi_5
-4\eta_3\eta_6\chi_4\chi_5+\cdots .
\end{align*}
The parts of \(Q_1\cdot V=Q_1^iV^i\) needed below are
\begin{align*}
Q_1\cdot V\supset{}&
4\eta_2\chi_1\chi_2
+4\eta_3\chi_1\chi_3
+4\eta_4\chi_2\chi_3\\
&+4\eta_5\chi_1\chi_5
+4\eta_6\chi_2\chi_5
+4\eta_7\chi_3\chi_5
+2\eta_8\chi_4\chi_5+\cdots .
\end{align*}
We look for a relation
\begin{equation}
\alpha S_1+\beta S_2+\gamma S_3+\delta S_4+\varepsilon S_5=0.
\label{eq:app-fierz-linear-combination}
\end{equation}

The coefficients are determined by expanding the five \(S_i\)'s in the ordered Grassmann monomial basis.

\subsection{First Monomial}

Take \(M_1=\eta_1\eta_2\eta_3\eta_4
\chi_1\chi_2\chi_3\chi_4\chi_5\). For \(S_1=Q_1^1\gamma^2\), since
\(Q_1^1=\chi_5\), we need the coefficient of
\(\eta_1\eta_2\eta_3\eta_4\chi_1\chi_2\chi_3\chi_4\) inside
\(\gamma^2\). The relevant pairings are
\begin{align*}
&(8\eta_1\eta_2\chi_1\chi_2)
(8\eta_3\eta_4\chi_3\chi_4),\\
&(8\eta_1\eta_3\chi_1\chi_3)
(8\eta_2\eta_4\chi_2\chi_4),\\
&(8\eta_1\eta_4\chi_2\chi_3)
(8\eta_2\eta_3\chi_1\chi_4),
\end{align*}

together with the reverse ordering of each product. Therefore
\begin{equation}
[S_1]_{M_1}=2(8\cdot8+8\cdot8+8\cdot8)=384.
\label{eq:app-M1-S1}
\end{equation}

For \(S_4=\gamma V^1(Q_1\cdot V)\), the contributing terms are
\begin{align*}
&(8\eta_2\eta_3\chi_1\chi_4)
(\eta_1\chi_5)
(4\eta_4\chi_2\chi_3),\\
&(8\eta_2\eta_4\chi_2\chi_4)
(\eta_1\chi_5)
(4\eta_3\chi_1\chi_3),\\
&(8\eta_3\eta_4\chi_3\chi_4)
(\eta_1\chi_5)
(4\eta_2\chi_1\chi_2).
\end{align*}

Each gives \(-32\) after reordering into canonical Grassmann order, so
\begin{equation}
[S_4]_{M_1}=-96.
\label{eq:app-M1-S4}
\end{equation}
Hence
\begin{equation}
384\alpha-96\delta=0,
\qquad
\delta=4\alpha.
\label{eq:app-M1-constraint}
\end{equation}

\subsection{Second Monomial}

Take \(M_2=\eta_2\eta_3\eta_6\eta_7
\chi_1\chi_2\chi_3\chi_4\chi_5\).
The nonzero contributions come from \(S_2\) and \(S_4\).
For \(S_2=Q_1^1\gamma(V^1)^2\), use \(Q_1^1=\chi_5\),
\(\gamma\supset 8\eta_2\eta_3\chi_1\chi_4\), and
\(V^1\supset \eta_6\chi_2+\eta_7\chi_3\). The two orderings of the two
\(V^1\) factors give
\begin{equation}
[S_2]_{M_2}=-16.
\label{eq:app-M2-S2}
\end{equation}
For \(S_4=\gamma V^1(Q_1\cdot V)\), the relevant products give
\begin{align*}
&(8\eta_2\eta_3\chi_1\chi_4)
(\eta_6\chi_2)
(4\eta_7\chi_3\chi_5)
\quad\Rightarrow\quad -32,\\
&(8\eta_2\eta_3\chi_1\chi_4)
(\eta_7\chi_3)
(4\eta_6\chi_2\chi_5)
\quad\Rightarrow\quad -32,\\
&(4\eta_2\eta_7\chi_4\chi_5)
(\eta_6\chi_2)
(4\eta_3\chi_1\chi_3)
\quad\Rightarrow\quad -16,\\
&(-4\eta_3\eta_6\chi_4\chi_5)
(\eta_7\chi_3)
(4\eta_2\chi_1\chi_2)
\quad\Rightarrow\quad -16.
\end{align*}
Therefore
\begin{equation}
[S_4]_{M_2}=-96.
\label{eq:app-M2-S4}
\end{equation}
Thus
\begin{equation}
-16\beta-96\delta=0,
\qquad
\beta=-24\alpha,
\label{eq:app-M2-constraint}
\end{equation}
where we used \(\delta=4\alpha\).

\subsection{Third Monomial}

Take \(M_3=\eta_1\eta_2\eta_3\eta_4
\chi_1\chi_2\chi_5\chi_7\chi_8\).
The nonzero contributions come from \(S_2\) and \(S_5\). For \(S_2\), use
\(Q_1^1=\chi_5\), \(\gamma\supset 8\eta_1\eta_2\chi_1\chi_2\), and
\(V^1\supset \eta_3\chi_7+\eta_4\chi_8\). The two orderings of the
\(V^1\) factors give
\begin{equation}
[S_2]_{M_3}=-16.
\label{eq:app-M3-S2}
\end{equation}
For
\(S_5=(V^1)^3(Q_1\cdot V)\), use
\(V^1\supset \eta_1\chi_5+\eta_3\chi_7+\eta_4\chi_8\) and
\(Q_1\cdot V\supset 4\eta_2\chi_1\chi_2\). There are \(3!\) orderings of
the three \(V^1\) factors, each contributing \(4\). Hence
\begin{equation}
[S_5]_{M_3}=24.
\label{eq:app-M3-S5}
\end{equation}
Therefore
\begin{equation}
-16\beta+24\varepsilon=0,
\qquad
\varepsilon=-16\alpha,
\label{eq:app-M3-constraint}
\end{equation}
where we used \(\beta=-24\alpha\).

\subsection{Fourth Monomial}

Take \(M_4=\eta_5\eta_6\eta_7\eta_8
\chi_1\chi_2\chi_3\chi_4\chi_5\).
The nonzero contributions come from \(S_3\) and \(S_5\). For
\(S_3=Q_1^1(V^1)^4\), use \(Q_1^1=\chi_5\) and
\[
V^1\supset
\eta_5\chi_1+\eta_6\chi_2+\eta_7\chi_3+\eta_8\chi_4.
\]
There are \(4!\) orderings, each contributing \(+1\), so $[S_3]_{M_4}=24$.
For \(S_5=(V^1)^3(Q_1\cdot V)\), the factor \(Q_1\cdot V\) can supply
\(4\eta_5\chi_1\chi_5\), \(4\eta_6\chi_2\chi_5\),
\(4\eta_7\chi_3\chi_5\), or \(2\eta_8\chi_4\chi_5\).
The remaining three \(\eta\chi\) pairs come from \(V^1\). The first three
choices give \(3!\cdot4=24\) each, while the last gives \(3!\cdot2=12\).
Therefore
\begin{equation}
[S_5]_{M_4}=24+24+24+12=84.
\label{eq:app-M4-S5}
\end{equation}
Thus
\begin{equation}
24\gamma+84\varepsilon=0,
\qquad
\gamma=56\alpha,
\label{eq:app-M4-constraint}
\end{equation}
where we used \(\varepsilon=-16\alpha\).

\subsection{Result}

We have found
\begin{equation}
\delta=4\alpha,
\qquad
\beta=-24\alpha,
\qquad
\varepsilon=-16\alpha,
\qquad
\gamma=56\alpha.
\label{eq:app-fierz-coefficients}
\end{equation}
Choosing \(\alpha=1\), the unique linear relation is
\begin{equation}
S_1-24S_2+56S_3+4S_4-16S_5=0.
\label{eq:app-fierz-linear-relation}
\end{equation}
Restoring the definitions of \(S_i\), this gives
\begin{equation}
Q_a^1\gamma^2
-24Q_a^1\gamma(V^1)^2
+56Q_a^1(V^1)^4
+4\gamma V^1(Q_a\cdot V)
-16(V^1)^3(Q_a\cdot V)
=0.
\label{eq:app-final-fierz-identity}
\end{equation}
This is the \(SO(8)\) Fierz identity used in the supersymmetry constraint.
\section{One-loop SYM (covariant form)}\label{app:sym-compact}

In this appendix we rewrite the one-loop calculation in a more covariant notation, suitable for other backgrounds.
The Euclidean action is
\begin{align}
S_E
&=
\frac{1}{g_{\rm YM}^2}
\int d^2\sigma\,
\mathrm{Tr}\left[
\frac14F_{\alpha\beta}F_{\alpha\beta}
+
\frac12D_\alpha X^iD_\alpha X^i
-
\frac14[X^i,X^j]^2
+\text{fermions}
\right],
\label{eq:canonical-D1-action}
\end{align}
where \(\alpha,\beta=0,1\) and \(i,j=1,\ldots,8\).  The spatial circle is
\(\sigma\sim \sigma+2\pi R\).

\subsection{Ten-dimensional notation}

We can package the two-dimensional gauge field and the eight scalars into a formal
ten-dimensional gauge field,
\begin{equation}
A_M=(A_\alpha,X^i),
\qquad
M,N=0,\ldots,9.
\end{equation}
The corresponding field strengths are
\begin{equation}
\mathcal F_{\alpha\beta}=F_{\alpha\beta},
\qquad
\mathcal F_{\alpha i}=D_\alpha X^i,
\qquad
\mathcal F_{ij}=i[X^i,X^j].
\end{equation}
For a slowly varying Coulomb-branch background, the common massive scalar
operator for one complex off-diagonal multiplet is
\begin{equation}
\mathcal O=k^2+r_X^2.
\end{equation}
where
\begin{equation}
r_X^2=\Delta X^i\Delta X^i.
\end{equation}

\subsection{Quadratic operators}

In background-field gauge, the quadratic bosonic operator is
\begin{equation}
\Delta^{\rm B}_{MN}
=
\mathcal O\,\delta_{MN}-2i\mathcal F_{MN}.
\end{equation}
The complex ghost pair sees
\begin{equation}
\Delta_{\rm gh}=\mathcal O.
\end{equation}
The squared fermion operator is
\begin{equation}
\Delta_{\rm F}^2
=
\mathcal O-\frac{i}{2}\Gamma^{MN}\mathcal F_{MN}.
\end{equation}
Thus the one-loop determinant is
\begin{equation}
\log Z_1
=
-\mathrm{Tr}_{10}\log(\mathcal O\delta_{MN}-2i\mathcal F_{MN})
+2\mathrm{Tr}\log\mathcal O
+\frac12\mathrm{Tr}_{16}\log\left(\mathcal O-\frac{i}{2}\Gamma^{MN}\mathcal F_{MN}\right).
\label{eq:oneloop-determinant-common}
\end{equation}
The three terms are the ten bosonic components, the complex ghosts, and the
sixteen-component Majorana--Weyl fermion.

\subsection{Expansion of the determinant}

For any shift matrix \(\Sigma\),
\begin{equation}
\mathrm{Tr}\log(\mathcal O+\Sigma)
=
\mathrm{Tr}\log\mathcal O
+
\sum_{n=1}^\infty
\frac{(-1)^{n+1}}{n}
\mathrm{Tr}\left[(\mathcal O^{-1}\Sigma)^n\right].
\end{equation}
Here
\begin{equation}
\Sigma_V{}^M{}_N=-2i\mathcal F^M{}_N,
\qquad
\Sigma_S=-\frac{i}{2}\Gamma^{MN}\mathcal F_{MN}.
\end{equation}
The zeroth-order supertrace cancels:
\begin{equation}
-10+2+\frac12(16)=0.
\end{equation}
The quadratic term cancels because
\begin{equation}
-\mathrm{tr}_{10}(\Sigma_V^2)
+
\frac12\mathrm{tr}_{16}(\Sigma_S^2)=0.
\end{equation}
Odd powers vanish by the trace over an odd number of antisymmetric Lorentz
generators, so the first nonzero local term is quartic in \(\mathcal F\).

\subsection{The one-loop quartic term}

The quartic term is
\begin{equation}
\Gamma_1^{(4)}
=
6\int d^2\sigma
\left[
\mathcal F^4-\frac14(\mathcal F^2)^2
\right]
\int\frac{d^2k}{(2\pi)^2}\frac{1}{(k^2+r_X^2)^4}.
\end{equation}
Equivalently,
\begin{align}
\Gamma_1^{(4)}
&=
6\int d^2\sigma
\left[
\mathcal F_{MN}\mathcal F^{NP}\mathcal F_{PQ}\mathcal F^{QM}
-
\frac14
\mathcal F_{MN}\mathcal F^{MN}\mathcal F_{PQ}\mathcal F^{PQ}
\right]
\nonumber\\
&\hspace{1cm}\times
\int\frac{d^2k}{(2\pi)^2}\frac{1}{(k^2+r_X^2)^4}.
\end{align}
The two-dimensional integral is
\begin{equation}
\int\frac{d^2k}{(2\pi)^2}\frac{1}{(k^2+r_X^2)^4}
=
\frac{1}{12\pi r_X^6}.
\end{equation}
Therefore
\begin{equation}
\Gamma_{1,{\rm D1}}^{(4)}
=
\frac{1}{2\pi r_X^6}
\int d^2\sigma
\left[
\mathcal F_{MN}\mathcal F^{NP}\mathcal F_{PQ}\mathcal F^{QM}
-
\frac14
\mathcal F_{MN}\mathcal F^{MN}\mathcal F_{PQ}\mathcal F^{PQ}
\right].
\label{eq:D1-compact-F4-common}
\end{equation}

\subsection{Reduction to the two-body background}

We now evaluate \eqref{eq:D1-compact-F4-common} on the same slowly moving
two-body Cartan background used in the main text.  Take the relative
separation in the ultraviolet SYM normalization to be
\begin{equation}
\Delta X^1=v_X\tau,
\qquad
\Delta X^2=b_X,
\qquad
\Delta X^{i}=0\quad (i\neq 1,2),
\label{eq:app-c-background}
\end{equation}
with vanishing background gauge field.  In the ten-dimensional notation this
means that the only nonzero component of the field strength is
\begin{equation}
\mathcal F_{\tau 1}
=
\partial_\tau \Delta X^1
=
v_X,
\qquad
r_X^2=b_X^2+v_X^2\tau^2 .
\label{eq:app-c-field-strength}
\end{equation}
For a single antisymmetric \(2\times2\) block with
\(\mathcal F_{\tau1}=v_X\), the two invariants are
\begin{equation}
\mathcal F_{MN}\mathcal F^{MN}=2v_X^2,
\qquad
\mathcal F_{MN}\mathcal F^{NP}\mathcal F_{PQ}\mathcal F^{QM}=2v_X^4.
\label{eq:app-c-invariants}
\end{equation}
Therefore the quartic combination appearing in the determinant reduces to
\begin{equation}
\mathcal F_{MN}\mathcal F^{NP}\mathcal F_{PQ}\mathcal F^{QM}
-
\frac14
\mathcal F_{MN}\mathcal F^{MN}\mathcal F_{PQ}\mathcal F^{PQ}
=
v_X^4 .
\label{eq:app-c-f4-reduction}
\end{equation}
For one off-diagonal complex multiplet, \eqref{eq:D1-compact-F4-common} gives
the local contribution
\begin{equation}
\Delta \mathcal L_{\rm 1-loop}^{(X)}
=
\frac{1}{2\pi}
\frac{v_X^4}{r_X^6}.
\label{eq:app-c-local-X}
\end{equation}
Integrating over the spatial circle and multiplying by the \(N_1N_2\)
off-diagonal sectors between the two clusters gives
\begin{equation}
\Delta L_{\rm 1-loop}
=
2\pi R\,N_1N_2\,
\Delta \mathcal L_{\rm 1-loop}^{(X)}
=
R\,N_1N_2\frac{v_X^4}{r_X^6}.
\label{eq:app-c-zero-mode-X}
\end{equation}
Finally use the normalization dictionary
\begin{equation}
X^i=\frac{g_{\rm YM}}{\sqrt{2\pi}}x^i,
\qquad
\frac{v_X^4}{r_X^6}
=
\frac{2\pi}{g_{\rm YM}^2}\frac{v^4}{r^6}.
\label{eq:app-c-X-to-x}
\end{equation}
Thus the covariant one-loop determinant reproduces the main-text result,
\begin{equation}
\Delta L_{\rm 1-loop}
=
2\pi R\,
\frac{N_1N_2}{g_{\rm YM}^2}
\frac{v^4}{r^6}.
\label{eq:app-c-main-answer}
\end{equation}
\bibliography{biblio}
\bibliographystyle{jhep.bst}
\end{document}